\documentclass[11pt,preprintnumbers,nofootinbib,prd]{revtex4-1}
\usepackage{graphicx}
\usepackage{amsmath}
\usepackage{epsfig}
\usepackage{mathrsfs}

\newcommand{\be}{\begin{equation}}
\newcommand{\ee}{\end{equation}}
\newcommand{\ba}{\begin{eqnarray}}
\newcommand{\ea}{\end{eqnarray}}

\begin{document}
\input{epsf}
\begin{flushright}
SLAC-PUB-13442
\end{flushright}

\title{Kahler Independence of the $G_2$-MSSM}
\author{Bobby S. Acharya}
\affiliation{Abdus Salam International Centre for Theoretical
Physics, Strada Costiera 11, Trieste, Italy\\and\\INFN, Sezione di Trieste}
\author{Konstantin Bobkov}
\affiliation{Stanford Linear Accelerator Center,\\ 2575 Sand Hill Road, Menlo Park, CA 94025}

\vspace{0.5cm}

\date{\today}

\vspace{0.3cm}

\begin{abstract}
The $G_2$-MSSM is a model of particle physics coupled to moduli fields with interesting
phenomenology both for colliders and astrophysical experiments. In this paper we consider
a more general model - whose moduli Kahler potential is a {\it completely arbitrary} 
$G_2$-holonomy Kahler potential
and whose matter Kahler potential is also more general. We prove that
the vacuum structure and spectrum of BSM particles is largely unchanged in this much more general
class of theories. In particular, gaugino masses are still suppressed relative to the gravitino mass and
moduli masses. We also consider the effects of higher order corrections to the matter Kahler potential
and find a connection between the nature of the LSP and flavor effects.

\end{abstract}
\maketitle
\newpage
\vspace{-1.2cm} \tableofcontents

\section{Introduction}

From several theoretical points of view, the existence of 
moduli fields seems inevitable. For instance,
supersymmetry may be the mechanism responsible for stabilizing the
scale of the Standard Model. Supersymmetry requires supergravity, whose
only (known) reasonable UV completion seems to be String theory; and along with
string theory come extra dimensions and their moduli. In fact, since string/$M$ theory
contains no dimensionless parameters, moduli appear necessary to explain the observed values
of various couplings in nature. From the bottom up, moduli appear in various theories
with "dynamical couplings" as well as in Inflation -- the inflaton field is usually a
neutral scalar field aka a modulus. For all of these reasons and more, moduli physics and
phenomena must be considered seriously. 

In a series of papers, \cite{Acharya:2007rc}, \cite{Acharya:2008zi}, \cite{Acharya:2008bk}, 
a very detailed model of moduli physics coupled to matter
has been described. The $G_2$-MSSM model, largely inspired by $M$ theory compactifications
on manifolds of $G_2$-holonomy, is a model in which strong gauge dynamics in the hidden
sector generates a potential which both stabilizes all the moduli fields and simultaneously
generates a hierarchically small scale -- thus solving (most of) the hierarchy problem.
The model has an interesting spectrum: moduli have masses in the 50-100 TeV region, scalar
superpartners and higgsinos have masses in the 10's of TeV region, whilst gauginos, which are
the lightest BSM particles have masses of order 100's of GeV. Direct production
of gluinos and electroweak gauginos are the dominant new physics channels at the LHC. The nature of the
LSP is also very interesting as it is a neutral Wino. Moreover, its production in the early universe
is dominated by the decays of the moduli fields (ie non-thermal production)
and can naturally account for the observed fraction of dark matter today.  
The moduli and gravitino problems are avoided due to the gravitino mass scale being 
one to two orders of magnitude larger than the TeV scale. One drawback of the model is the fine tuning
between the 10's of TeV scale and $M_Z$ and is the reason the model solves most of the
hierarchy problem and not all of it.

However, the $G_2$-MSSM model, as defined in \cite{Acharya:2008zi}, 
is based on some specific assumptions about the
moduli and matter Kahler potentials, albeit with the claim that these are
general enough to incorporate all of the essential ingredients of more general Kahler potentials (and hence
$G_2$-manifolds). Thus far, there has been no serious study of these assumptions and it is the main
aim of this paper to undertake this. The main result that we prove here is that the mass spectrum of the theory 
depends very weakly on the specific form of the {\it moduli} Kahler potential; in fact the 
spectrum depends
on the Kahler potential for moduli {\it only} through the fact that it is the Log of a homogeneous function
(the volume of the extra dimensions); the precise nature of this homogeneous function is fairly irrelevant as we
will see. 

We also discuss the Kahler potential for {\it charged matter fields}. We give three consistent arguments for calculating the
moduli dependence of the matter kinetic terms in 4d Einstein frame. Whilst non-trivial, these modifications do
not change the results of  \cite{Acharya:2007rc}, \cite{Acharya:2008zi} much. 
More importantly, we also consider higher order terms in the matter Kahler potential,
in particular the terms which are usually considered troublesome for flavor physics in theories of
gravity mediated susy breaking. Whilst we expect that such operators will be suppressed, if they enter with
large coefficients they can affect the mass spectrum: 1) they can directly alter the scalar and higgsino masses, 
which are typically large; 2) they can indirectly (via threshold effects from the higgsinos) alter the nature of the LSP.
In particular, we find that the LSP can also be a Bino in some cases. This provides
a connection between flavor physics and the nature of the LSP in models of this sort.

The paper is organized as follows. The next section describes some simple properties of
the moduli space metric for general $G_2$-manifolds which will be important for our
later considerations. Section III is devoted to the Kahler potential for charged
matter fields. Following this, we re-do the analysis of Moduli stabilization from \cite{Acharya:2007rc}
in this much more general context. In section V we compute the mass spectrum and susy breaking
couplings in the minimum of the potential and demonstrate that it is almost identical to that
of the original $G_2$-MSSM. In section VI we present a further generalization of the construction.
In section VII we renormalize the Lagrangian down to the Electroweak scale and give the spectrum there.

\section{General properties of moduli space metrics on $G_2$ holonomy manifolds}

In this section we describe some very general and simple properties of the moduli space
metric of $G_2$-manifolds. It is these simple properties, which will allow us to draw very
general conclusions.

The metric $g(X)$ on a $G_2$ holonomy manifold $X$ can be expressed in terms of the associative three-form $\Phi$ as
\be
g_{ij}=\left({\rm det}s\right)^{-\frac 19}s_{ij}
\ee
with
\be
s_{ij}=\frac 1{144}\Phi_{ikl}\Phi_{jnm}\Phi_{rst}{\hat \epsilon}^{\,klnmrst}\,,\,\,\,\,\,{\hat\epsilon}^{\,12...7}=+1.
\ee
Expanding $\Phi$ in terms of basis harmonic three-forms $\phi_i\in H^3(X,{\bf Z})$  (modulo torsion) we obtain
\be\label{phiexp}
\Phi=\sum_{i=1}^{N}s_i\phi_i\,,\,\,\,\,\,N=b^3(X)= dim\left({\cal M}(X)\right)\,,
\ee
where $s_i$ are geometric moduli corresponding to the perturbations of the internal metric.
The complexified moduli space ${\cal M}(X)$ of a $G_2$ holonomy compactification manifold $X$ has holomorphic coordinates $z_i$ given by
\be
z_i=t_i+is_i\,,
\ee
where $t_i$ are the axions parameterizing the zero modes of the 11-dimensional supergravity 
three-form $C_3$.
The classical moduli space metric (not including possible quantum corrections) 
can be derived from the following Kahler potential \cite{Beasley:2002db}
\be
\hat K= -3 \ln 4\pi^{1/3} V_X\,,
\ee
where the dimensionless volume $V_X\equiv Vol(X)/l_M^7$ is a homogeneous function of $s_i$ of degree $7/3$ and $l_M$
is the 11d Planck length. The homogeneity of $V_X$ is the key property that we will utilize in what follows.
In terms of the associative three-form $\Phi$, the volume is given by \cite{Beasley:2002db}
\be\label{vphi}
V_X=\frac 1 7\int_X\Phi\wedge *\Phi\,.
\ee
Define the following derivatives with respect to the moduli
\be
\hat K_i\equiv\frac{\partial \hat K}{\partial s_i}\,\,\,\,\,{\rm and }\,\,\,\,\hat K_{ij}\equiv\frac{\partial^2\hat K}{\partial s_i\partial s_j}\,.
\ee

The matrix $\hat K_{ij}$, the Hessian of $\hat K$ is related to the actual Kahler metric $\hat G_{i\bar j}$ which controls the kinetic terms as 
$4\hat G_{i\bar j}=\hat K_{i \bar j}$, where in the Hessian
we simply replace index $j$ with $\bar j$. 
Since $V_X$ is a homogeneous function of degree $7/3$, the first
derivative of $\hat K$ defined above has the following property
\be\label{eq1}
\sum_{i=1}^Ns_i\hat K_i=-7\,.
\ee
Differentiating (\ref{eq1}) with respect to $s_j$ we obtain an important
property of the metric $\hat K_{ij}$ 
\be\label{eq2}
\sum_{i=1}^Ns_i\hat K_{ij}=-\hat K_j\,,\,\,{\rm and \,\,since \,\,}\hat K_{ij}\,\,{\rm is\,\,symmetric\,\,}\sum_{j=1}^Ns_j\hat K_{ij}=-\hat K_i\,.
\ee
Let us now introduce a set of dual coordinates $\{\tau_i\}$ defined by
\be\label{eq14}
\tau_i\equiv \frac{\partial V_X}{\partial s_i}\,.
\ee
Note that the variables $\{\tau_i\}$ are homogeneous functions of $\{s_i\}$ of degree $4/3$ . 
Using the homogeneity of the volume together with the definition (\ref{eq14}) we can express the volume $V_X$ as
\be\label{v73}
\frac 73 V_X=\sum_{i=1}^N s_i \frac{\partial V_X}{\partial s_i}=\sum_{i=1}^N s_i \tau_i\,,\,\,\,\Rightarrow\,\,\,V_X=\frac 37\sum_{i=1}^N s_i \tau_i\,,
\ee
and 
\be
\hat K_i=\frac{\partial \hat K}{\partial s_i}=-\frac 3{V_X}\frac{\partial V_X}{\partial s_i}=-\frac{3\tau_i}{V_X}\,.
\ee

Combining (\ref{phiexp}),  (\ref{vphi}) and (\ref{v73}) we can reexpress the dual variables as
\be
\tau_i= \frac 13\int_X \phi_i\wedge *\Phi=\frac 13\int_{[\tau_i]}*\Phi\,,
\ee
where for each harmonic basis three-form $\phi_i\in H^3(X, {\bf Z})$ we introduced a Poincare dual four-cycle $[\tau_i]\in H_4(X)$.
We now use the above duality to make a particularly convenient choice of the basis harmonic three-forms.
In particular, we choose a basis $\{\phi_i\}\in H^3(X)$ such that the periods of the fundamental co-associative four-form $*\Phi$ 
over the Poincare dual four-cycles are positive definite:
\be\label{basischoice}
\int_{[\tau_i]}*\Phi>0\,.
\ee
This choice of a basis becomes obvious when we recall that for a generic basis four-cycle $[\tau_i]\in H_4(X)$
\be
Vol([\tau_i])\geq \frac 13\int_{[\tau_i]}*\Phi=\tau_i\,,
\ee
where the above relation becomes an equality if and only if the corresponding four-cycle is co-associative.

All geometric moduli describing the fluctuations of the internal metric must be massive in order to satisfy constraints from fifth force experiments and cosmology. At the same
time the vacuum expectation values of the moduli (coordinates $s_i$ on the moduli space) must be fixed in the region of the moduli space where the geometric 
description makes sense. 
In Section \ref{s3} we describe a way to stabilize the moduli, which ensures that we find isolated minima that satisfy these conditions automatically.

It turns out that for our purposes it is convenient to introduce a set of "angular" variables ${a}_i$ defined by
\be\label{eq3}
{a}_i\equiv -\frac 13 s_i\hat K_i=\frac{s_i\tau_i}{V_X}\,,\,\,{\rm no\,\,sum\,\,over\,\,i}\,.
\ee
We see that ${a}_i$ are scale-independent and satisfy
\be\label{eq4}
\sum_{i=1}^N{a}_i=\frac 73\,.
\ee
Thus, we can also parameterize the moduli space ${\cal M}(X)$ by a subset of $N-1$ variables $a_i$ plus one volume, e.g. the volume of the manifold $V_X$. 
Differentiating the ${a}_i$ allows us to introduce the matrix
\be\label{eq5}
P_{ij}\equiv -s_j\frac{\partial {a}_i}{\partial s_j}\,,\,\,{\rm no\,\,sum\,\,over\,\,j}\,.
\ee
which has components
\be\label{eq6}
P_{ij}= \frac 13 \delta_{ij}s_j\hat K_i+s_is_j\frac 13\hat K_{ij}\,,\,\,{\rm no\,\,sum\,\,over\,\,i,\,j}\,.
\ee
$P_{ij}$ has the following contraction properties, which follow from (\ref{eq4}) and the fact that $a_i$ are homogeneous of degree zero
\be\label{eq7}
\sum_{i=1}^NP_{ij}=0\,,\,\,\,{\rm and}\,\,\,\sum_{j=1}^NP_{ij}=0\,.
\ee
We can then write
\be\label{eq9}
\hat K_{ij}=\frac{3{a}_j}{s_is_j}\Delta_{ij}\,, 
\ee
where the matrix $\Delta_{ij}$ is defined as
\be\label{eq10}
\Delta_{ij}\equiv \delta_{ij}+\frac{P_{ij}}{{a}_j}\,,
\ee
and satisfies the following contraction properties
\be\label{eq11}
\sum_{i=1}^N\Delta_{ij} =1\,,\,{\rm and}\,\,\,\sum_{j=1}^N\Delta_{ij}{a}_j ={a}_i\,,
\ee
where we used (\ref{eq7}) to derive (\ref{eq11}). Note that parameters $a_i$ defined 
in (\ref{eq3}) are the components of an eigenvector $a$
of the non-Hermitian matrix $\Delta$ with unit eigenvalue.

We can compute the formal inverse of the Hessian metric, $\hat K^{ij}$. By definition of the inverse it must satisfy
\be
\sum_{j=1}^N\hat K^{ij}\hat K_{jk}=\delta^i_k\,,
\ee
and using (\ref{eq9}) it can be expressed as
\be\label{eq12}
\hat K^{ij}=\frac{s_is_j}{3{a}_i}(\Delta^{-1})^{ij}\,,
\ee
where the inverse matrix $(\Delta^{-1})^{ij}$ satisfies
\be\label{eq13}
\sum_{j=1}^N(\Delta^{-1})^{ij}\Delta_{jk}=\delta^i_k\,.
\ee
Symbolically we can express $\Delta^{-1}$ as
\be
\Delta^{-1}=\frac 1{1+\frac P{a}}\,,
\ee
which in terms of components translates into
\be\label{InverseD}
(\Delta^{-1})^{ij}=\delta^{ij}-P_{ij}\frac 1{{a}_j}
+{P_{il}}\frac 1{{a}_l}{P_{lj}}\frac 1{{a}_j}
-{P_{il}}\frac 1{{a}_l}{P_{lm}}\frac 1{{a}_m}{P_{mj}}\frac 1{{a}_j}+\,.\,.\,.\,.
\ee
Using (\ref{eq7}) and (\ref{InverseD}) we derive the following properties of the inverse matrix
$\Delta^{-1}$
\be\label{contrD}
\sum_{i=1}^N(\Delta^{-1})^{ij} =1\,,\,{\rm and}\,\,\,\sum_{j=1}^N(\Delta^{-1})^{ij}{a}_j ={a}_i\,,
\ee
which could have also been obtained directly from (\ref{eq11}).
Note that although we do not have a closed form expression for the components 
$(\Delta^{-1})^{ij}$, the contraction properties in (\ref{contrD}) are what will ultimately
allow us
to derive explicit expressions for the terms in the soft breaking lagrangian -- since such
couplings depend only on the contractions and not the precise details of the
functional form of $V_X$. Before going on to the details of these calculations, we first
must consider the Kahler potential for matter fields in $M$ theory.

\section{Kahler potential for charged chiral matter}

In this section we re-visit the Kahler potential for charged matter fields in $M$ theory.
In practice, the absence of a useful microscopic formulation
makes it difficult to compute the moduli dependence
of the Kahler potential for these fields in general. Below we outline three arguments
for the structure of the Kahler potential - first from dimensional reduction,
second based on the scaling properties of physical Yukawa couplings
and the third based on the form of the threshold corrections to the physical gauge coupling. 
Happily, all three methods agree.

\subsection{Kahler potential from dimensional reduction}

In $M$ theory, charged chiral matter is localized near conical singularities 
\cite{Atiyah:2001qf}, \cite{Witten:2001uq}, \cite{Acharya:2001gy}, \cite{Acharya:2004qe}.
These are literally points in the seven extra dimensions. Because of this,
we expect that the kinetic terms for the chiral matter fields should be
"largely independent of bulk moduli fields" that the $G_2$ manifold X has.
The precise meaning of this statement will be clarified below in terms of the
scaling property of the kinetic term.
They could, of course depend on local moduli inherent to the conical singularity,
but, since, in a supersymmetric theory, a single chiral multiplet in a complex
representation of the gauge group usually has no $D$ or $F$-flat directions
\cite{Luty:1995sd}, there are
typically no such local moduli.

There is a subtlety in the above general arguments. Since, in four dimensions,
a scalar field kinetic term is not invariant under Weyl rescalings of the metric,
one has to pick a Weyl gauge. We will argue that the correct Weyl gauge for the statement
above is NOT the 4d Einstein frame. Therefore, the kinetic term for chiral matter will be
non-trivial in the 4d Einstein frame, which is the standard one in which to define the Kahler potential.

Since the physics of a conical singularity in $M$ theory does not introduce any
new scale, asides from the 11d Planck scale, the only reasonable Weyl frame is the
11d Einstein frame.  Therefore the lagrangian density in the 11d frame is
\be\label{lb1}
L \sim M_{11}^9 \sqrt{g_{11}} R + \delta_7 \wedge  \partial_{M} \phi \partial_N \phi^{\dagger} g^{MN}\kappa(s_i) \sqrt{g_{11}} +\,.\,.\,.\,,
\ee
where $\delta_7$ is a delta function peaked at the position of the matter multiplet containing the scalar field $\phi$ and has mass dimension seven. $\kappa(s_i)$
is a homogeneous function of the moduli of degree zero which will generally be of order one and vary adiabatically.\footnote{In the toroidal Type IIA compactifications with intersectiong D6-branes, the analog of $\kappa$ is a scale invariant 
function that depends on the relative intersection angles $\theta_i^{\alpha}$ \cite{Blumenhagen:2007ip}.}, i.e.
\be\label{hpr2}
\sum_{k=1}^Ns_k\frac{\partial \kappa(s_i)}{\partial s_k}=0\,.
\ee

The above property implies that $\kappa(s_i)$ remains invariant when the moduli are rescaled as $s_i\rightarrow\lambda s_i$,
thus explicitly implementing the idea that in the 11d frame the kinetic term of a matter field localized at a point $p\in X$
is "largely independent of bulk moduli". A particularly simple example satisfying (\ref{hpr2}) is when $\kappa(s_i)=const$.
Integrating this over $X$ leads to a 4d density
\be
L_4 \sim V_X M_{11}^2 \sqrt{g_4} R_4 + \kappa(s_i) g_4^{\mu\nu} \partial_{\mu} \phi \partial_{\nu} \phi^{\dagger}  \sqrt{g_4}\,,
\ee
where $V_X$ is the volume of the extra dimensions in 11d units. This is the Lagrangian in 11d Einstein frame. 
If we now Weyl rescale into the 4d Einstein frame we find
\be
L_4 \sim \frac 1{2\kappa_4^2}\sqrt{g_E} R_E + {\frac {\kappa(s_i)} {V_X}}g_E^{\mu\nu} \partial_{\mu} \phi \partial_{\nu} \phi^{\dagger}  \sqrt{g_E}\,,
\ee
where the subscript $E$ indicates that we are using the 4d Einstein frame metric.

We have only considered the Einstein-Hilbert and kinetic terms of the matter fields. Including all the other
terms would give the 4d supergravity Lagrangian in Einstein frame. In particular, from this we would read off
that the Kahler metric for the multiplet containing $\phi$ is
\be\label{kmet}
{\tilde K}_{\phi\bar\phi} = {\frac {\kappa(s_i)} {V_X}}\,.
\ee
If we introduce dimensionless fields $\hat\phi$ as in $\phi=m_{pl}\hat\phi$, the Kahler potential is
\be
{\tilde K}=\kappa(s_i)\frac{\phi\phi^{\dagger}}{V_X}=\kappa(s_i)\frac{\hat\phi{\hat\phi}^{\dagger}}{V_X}m_{pl}^2\,.
\ee
As we will see, this is consistent with the arguments given in the next subsection.

\subsection{Kahler metric from the properties of the physical Yukawa couplings}

Here we will describe an alternative way of deducing the volume dependence of the Kahler metric
for charged chiral matter. This method is due to Conlon, Cremades and Quevedo \cite{Conlon:2006tj} and utilizes the relation 
between the physical (normalized) Yukawa couplings $Y_{\alpha\beta\gamma}$ and the unnormalized Yukawa couplings 
$Y_{\alpha\beta\gamma}^{\prime}$ that appear in the supergravity superpotential.
Recall that in $G_2$ compactifications of $M$ theory, a superpotential Yukawa coupling $Y_{\alpha\beta\gamma}^{\prime}$
between the multiplets $\alpha$, $\beta$, $\gamma$ that are
localized at three co-dimension seven singularities is induced by an M2-brane instanton wrapping a supersymmetric three-cycle connecting the
three singular points. The absolute value of the Yukawa coupling is given by
\be
|Y_{\alpha\beta\gamma}^{\prime}|\sim e^{-2\pi V_{\alpha\beta\gamma}}\,, 
\ee
where
\be
V_{\alpha\beta\gamma}=\sum_im_i^{\alpha\beta\gamma}s_i\,,
\ee
is the volume of the supersymmeric three-cycle.
After we diagonalize the Kahler metric for the matter fields and go to the canonical basis,
the relation between the absolute values of the physical and unnormalized Yukawa couplings is simply 
a rescaling by $e^{K/2}\left(\tilde K_{\alpha}\tilde K_{\beta}\tilde K_{\gamma}\right)^{-1/2}$
\be\label{Yt}
\left|Y_{\alpha\beta\gamma}\right|=e^{K/2}\left|Y_{\alpha\beta\gamma}^{\prime}\right|
\left(\tilde K_{\alpha} \tilde K_{\beta}\tilde K_{\gamma}\right)^{-1/2}\sim\left|Y_{\alpha\beta\gamma}^{\prime}\right|
\left(V_X^3\tilde K_{\alpha} \tilde K_{\beta}\tilde K_{\gamma}\right)^{-1/2}\,.
\ee

On the other hand, one can construct perfectly well-defined seven-dimensional local models where the $G_2$ manifold 
is non-compact, e.g. an ALE-fibration over a three-sphere or a quotient thereof, in which case 
$V_X\rightarrow\infty\Rightarrow m_{pl}/M_{11}\rightarrow\infty$ and gravity is effectively decoupled.
Such models can also contain charged chiral matter fields and since their interactions are determined locally,
the corresponding physical Yukawa couplings should not vanish when gravity is decoupled.
Local models of this type can be obtained via lifting effective theories on intersecting 
D6-branes in Type IIA to $M$ theory.

Therefore, locality implies that the physical Yukawa couplings should be independent of the overall volume $V_X$
in the limit $V_X\rightarrow\infty$. For that to happen, the Kahler metrics ${\tilde K}_{\alpha}$, ${\tilde K}_{\beta}$, 
${\tilde K}_{\gamma}$ in (\ref{Yt}) must scale with the volume $V_X$ as
\be
{\tilde K}_{\alpha}\sim{\tilde K}_{\beta}\sim{\tilde K}_{\gamma}\sim\frac 1{V_X}\,,
\ee
which is in perfect agreement with the form of the Kahler metric derived in the previous subsection.

\subsection{Consistency check for the Kahler metric}
In this section we confirm the form of the Kahler metric for charged chiral matter by comparing 
the threshold corrections to the physical gauge couplings in $G_2$ compactifications 
of $M$ theory with the general results in ${\cal N}=1$ $D=4$ supergravity.

Let us first consider a hidden sector containing a pure glue $SU(N)$ supersymmetric Yang-Mills theory.
Using the notation in \cite{Friedmann:2002ty} we have the following relation for the gauge coupling at one loop
\be
\frac {16\pi^2}{g^2(\mu)}=\frac{16\pi^2}{g^2_M}-3N\ln\left(\frac {\Lambda^2}{\mu^2}\right)+{\cal S}\,,
\ee
where ${\cal S}$ are the one-loop threshold corrections, $g(\mu)$ is the physical gauge coupling
and $g_M$ is the tree-level Wilsonian gauge coupling. In our convention $g_M$ is related to the gauge kinetic function $f$ as
\be
\frac{4\pi}{g^2_M}={\rm Im}f\,.
\ee
Recall that the Wilsonian gauge coupling gets renormalized at one loop only. On the other hand
the physical coupling $g(\mu)$ is renormalized to all orders. The threshold corrections 
come from massive states and are independent of the scale $\mu$. Based on the topological arguments \cite{Friedmann:2002ty}, 
the threshold corrections due to Kaluza Klein modes in $G_2$ compactifications of $M$ theory are rather simple 
and can be calculated even without knowing the $G_2$ metric!
Friedmann and Witten \cite{Friedmann:2002ty} explicitly computed one-loop threshold corrections
due to the heavy Kaluza-Klein modes living on a supersymmetric cycle ${\cal Q}$ with 
$b_1({\cal Q})=b_2({\cal Q})=0$ and a non-trivial fundamental group. Such corrections come in a form of linear combinations
of Ray-Singer analytic torsions \cite{Ray}, which are topological invariants of ${\cal Q}$. For the case at hand, the threshold 
corrections are given by
\be
{\cal S}=2N\ln V_{\cal Q}\Lambda^3+2\sum_{i}{\cal T}_i{\rm Tr}_{{\cal R}_i}Q^2\,,
\ee
where $V_{\cal Q}$ is the volume of the supersymetric cycle ${\cal Q}$, ${\cal T}_i$ are the Ray-Singer 
torsions corresponding to different irreducible representations of the fundamental group
and $Q$ are the generators of $SU(N)$. Here, the cutoff dependence appears as 
a correction due to the zero mode contributions transforming in the trivial representation of 
$\pi_1(\cal Q)$. Once the threshold corrections are included explicitly, a somewhat unexpected cancellation 
of the $\Lambda$-dependence occurs \cite{Friedmann:2002ty} and the one-loop relation can be written as
\be
\frac {16\pi^2}{g^2(\mu)}=\frac{16\pi^2}{g^2_M}-3N\ln\left(\frac 1 {V_{\cal Q}^{2/3}\mu^2}\right)+{\cal S}^{\prime}
\ee
where 
\be
{\cal S}^{\prime}=2\sum_{i}{\cal T}_i{\rm Tr}_{{\cal R}_i}Q^2\,.
\ee
In fact, the cancellation of the $\Lambda$-dependence occurs for any supersymmetric cycle ${\cal Q}$
with $b_1({\cal Q})=b_2({\cal Q})=0$ \cite{Friedmann:2002ty}.

Now we would like to consider a more general case when the gauge theory is a supersymmetric QCD
with $N_f$ flavors of chiral matter fields $Q_{\alpha}$ transforming in $N$ of $SU(N)$ plus 
$N_f$ flavors of $\tilde Q_{\alpha}$ transforming in $\overline N$. Each chiral matter field transforming
in a complex representation arises from a separate co-dimension seven conical 
singularity on $X$ with each singular point $P_i\in{\cal Q}$. It was argued in \cite{Friedmann:2002ty} that
the singularities producing charged chiral matter fields have no effect on the KK harmonics
of the seven-dimensional vector multiplet. Moreover, since the conical singularities
introduce no new scale below the eleven-dimensional Planck scale $M_{11}$, the effective cutoff scale for these 
multiplets is naturally $M_{11}$. Including such multiplets into the running is straightforward and results in
\be\label{1loop}
\frac {16\pi^2}{g^2(\mu)}=\frac{16\pi^2}{g^2_M}-3N\ln\left(\frac 1 {V_{\cal Q}^{2/3}\mu^2}\right)+N_f\ln\left(\frac {M_{11}^2} {\mu^2}\right)+{\cal S}^{\prime}\,.
\ee
In addition to the KK thresholds, there may be some unknown corrections due to possible charged massive matter fields
with masses of order $M_{11}$. At this point we cannot say with certainty whether such massive charged $M$ theory modes 
are present in the spectrum but we cannot exclude this possibility either. Just like the KK thresholds, these 
corrections cannot be holomorphic functions of the chiral 
multiplets $z_i$ describing the moduli of $X$ since the axion partners of the geometric moduli decouple from the 
computations of the threshold corrections. However, there may be some non-holomorphic as well as constant contributions 
from such massive charged states. For now we will simply assume that they are constant and result in a slight
shift of the tree-level gauge coupling. On the other hand, moduli dependent contributions may arise from non-perturbative 
corrections due to membrane instantons but they will be exponentially suppressed and can be safely neglected.

Our next task is to independently verify that the Kahler metric for the charged chiral matter fields
matches the previously obtained result (\ref{kmet}). Here we will use a strategy similar to the one in 
\cite{Billo:2007py}, \cite{Blumenhagen:2007ip} and compare (\ref{1loop}) with the corresponding one-loop expression in 
${\cal N}=1$ $D=4$ supergravity given by \cite{Gaillard:1992bt}
\be\label{1loopsugra}
\frac {16\pi^2}{g^2(\mu)}=\frac{16\pi^2}{g^2_M}-\left(3N-N_f\right)\ln\left(\frac {m_{pl}^2} {\mu^2}\right)
-\left(N-N_f\right)\hat K+2N\ln\left(\frac 1{g^2_M}\right)-2N_f\ln\left({\tilde K}_{\alpha\bar\alpha}\right)\,.
\ee
In the above expression, $\hat K=-3\ln 4\pi^{1/3} V_X$ is the Kahler potential for the moduli and ${\tilde K}_{\alpha\bar\alpha}$ is the
Kahler metric for the charged chiral matter fields. We can use the definition of the four-dimensional Newton's 
constant $\kappa_4=\sqrt{8\pi G_N}=1/m_{pl}$ in terms of the eleven-dimensional gravitational coupling $\kappa_{11}$
\be
\kappa_4^2\equiv\frac{\kappa_{11}^2}{V_Xl_M^7}\,,
\ee
in combination with the common convention $2\kappa_{11}^2=(2\pi)^8M_{11}^{-9}$ and $M_{11}=2\pi/l_M$ to obtain
\be
M_{11}^2=\frac{\pi m_{pl}^2}{V_X}\,.
\ee
Using the above relations together with $4\pi/g_M^2=V_{\cal Q}/l_M^3$ we have from (\ref{1loopsugra})
\ba\label{1loopsugra2}
\frac {16\pi^2}{g^2(\mu)}&=&\frac{16\pi^2}{g^2_M}-\left(3N-N_f\right)\ln\left(\frac {V_XM_{11}^2} {\pi\mu^2}\right)
+3\left(N-N_f\right)\ln 4\pi^{1/3} V_X+2N\ln\left(\frac {V_{\cal Q}}{4\pi l_M^3}\right)-2N_f\ln\left({\tilde K}_{\alpha\bar\alpha}\right)\,\nonumber\\
&=&\frac{16\pi^2}{g^2_M}-\left(3N-N_f\right)\ln\left(\frac {M_{11}^2} { 4\pi^{4/3}\mu^2}\right)
-2N_f\ln 4\pi^{1/3}V_X+2N\ln\left(\frac {V_{\cal Q}}{4\pi l_M^3}\right)-2N_f\ln\left({\tilde K}_{\alpha\bar\alpha}\right)\,\nonumber\\
&=&\frac{16\pi^2}{g^2_M}-3N\ln\left(\frac {1} {V_{\cal Q}^{2/3}\mu^2}\right)
+N_f\ln\left(\frac {M_{11}^2} {\mu^2}\right)-2N_f\ln\left(V_X {\tilde K}_{\alpha\bar\alpha}\right)-\ln\left({(2\pi)^{4N} (8\pi)^{2N_f}}\right)\,.
\ea
The appearance of the last term is most likely due to the convention used to define $M_{11}$ in terms of $\kappa_{11}$ as well as the ambiguity in
defining the relation between $l_M$ and $M_{11}$. Thus, we shall regard this term as an artifact and ignore it in further
discussion.
Comparing (\ref{1loopsugra2}) with the expression on the right hand side in (\ref{1loop})
we conclude that up to a constant multiplicative factor, Kahler metric for the charged chiral matter fields $Q_{\alpha}$ is
\be
{\tilde K}_{\alpha\bar\alpha}\sim\frac 1{V_X}\,,
\ee
which precisely matches the result obtained in the previous subsections. On the other hand, the constant term ${\cal S}^{\prime}$ 
in (\ref{1loop}) has no corresponding analog in (\ref{1loopsugra2}) and represents a genuine threshold correction
to the Wilsonian gauge coupling $g_M$.

In the framework of ${\cal N}=1$ $D=4$ supergravity, the RG-invariant scale where super QCD with $N>N_f$ 
becomes strongly coupled is 
\be\label{StrongScale}
\Lambda^{3N-N_f}=m_{pl}^{3N-N_f}e^{-\frac{8\pi^2}{g_M^2}-\frac {S^{\prime}}2}e^{-(N-N_f)\frac{\hat K}2}\,,
\ee
where the second exponential factor is due to the local SUSY.
The Affleck-Dine-Seiberg effective superpotential \cite{Affleck:1983mk} $W$ should be identified with
\be
e^{\hat K/2}W=\frac{(N-N_f)\tilde\Lambda^{\frac{3N-N_f}{N-N_f}}}{{\rm det}(Q\tilde Q)^{\frac 1 {N-N_f}}}\,,
\ee
where the gauge coupling inside $\tilde \Lambda$ is complexified.
Using (\ref{StrongScale}), up to an overall numerical constant we obtain
\be\label{sug}
W\sim(N-N_f){m_{pl}^{\frac{3N-N_f}{N-N_f}}}{\rm det}(Q\tilde Q)^{-\frac {1} {N-N_f}}e^{i\frac{2\pi}{N-N_f}f}e^{-\frac{S^{\prime}_a}{2(N-N_f)}}\,.
\ee
In (\ref{sug}), the dimensionful chiral matter fields\footnote{Here we suppressed the flavor index.} $Q$ can be
expressed in terms of dimensionless fields $\hat Q$ as
\be
Q=m_{pl}\hat Q\,,\,\,\,\,\,{\rm and}\,\,\,\,\,\tilde Q=m_{pl}\hat {\tilde Q}\,.
\ee
Then, the superpotential becomes
\be\label{spot1}
W=\tilde C(N-N_f)\,m_{pl}^3{\rm det}(\hat Q\hat{\tilde Q})^{-\frac {1} {N-N_f}}e^{i\frac{2\pi}{N-N_f}f}
e^{-\frac{S^{\prime}_a}{2(N-N_f)}}\,,
\ee
where $\tilde C$ is an overall numerical constant. In our further notation, we also define the following
constants
\be\label{constants}
C\equiv \tilde C\,e^{-\frac{S^{\prime}_a}{2(N-N_f)}}\,,\,\,\,{\rm and}\,\,\,\,\,\,A\equiv (N-N_f)\,C\,.
\ee
Let us now consider the case of $N_f=1$ flavors. Introducing an effective meson degree of freedom
\be\label{mesondef}
\phi\equiv\sqrt{2\hat Q\hat{\tilde Q}}\,,
\ee
we can rewrite the superpotential in terms of $\phi$ as
\be
W=A\,m_{pl}^3\phi^{-\frac {2} {N-1}}e^{i\frac{2\pi}{N-1}f}\,,
\ee
where we have absorbed the factor of $2^{1/(N-1)}$ into the normalization constant $C$.
Along the D-flat direction we have $\hat Q=\hat{\tilde Q}$ and the Kahler potential
for the matter fields can be rewritten in terms of the effective meson fields $\phi$ as
\be\label{meskp}
\tilde K=\kappa(s_i)\frac{\hat Q^{\dagger}\hat Q}{V_X}m_{pl}^2+\kappa(s_i)\frac{\hat{\tilde Q}^{\dagger}\hat{\tilde Q}}{V_X}m_{pl}^2
=\kappa(s_i)\frac {\bar\phi\phi}{V_X}m_{pl}^2\,.
\ee

\subsection{Higher order terms}

Based on three independent arguments we have been able to deduce the volume dependence of the Kahler metric
for charged chiral matter fields localized at co-dimension seven singularities.
Denoting the visible sector charged chiral matter fields by $Q^{\alpha}$ their Kahler potential is then
given by
\be\label{mkpt}
{\tilde K}=\frac{{\kappa}_{\alpha\bar\beta}(s_i)Q^{\alpha}Q^{\dagger\bar\beta}}{V_X}\,.
\ee
In the regime where the size of the supersymmetric cycle supporting the visible sector is large (this assumption is justified in the context of the MSSM where 
the corresponding volume is $\alpha_{GUT}^{-1}\approx 25$ ) we can perform a systematic expansion of ${\kappa}_{\alpha\bar\beta}(s_i)$ in the inverse volume of the cycle (weak coupling) so that in the leading order ${\kappa}_{\alpha\bar\beta}(s_i)$ is a homogeneous function of $s_i$ of degree $\lambda$, satisfying
\be\label{conp}
\sum_{i=1}^Ns_i\frac{\partial{\kappa}_{\alpha\bar\beta}(s_i)}{\partial s_i}=\lambda{\kappa}_{\alpha\bar\beta}(s_i)\,.
\ee
 
Based on the property that  a given charged chiral matter multiplet is localized at a point $p\in X$ we expect that 
$\lambda=0$, i.e. ${\kappa}_{\alpha\bar\beta}(s_i)$ is scale invariant in the leading order. 
Therefore, when the moduli are simultaneously scaled up by an overall positive constant, the  
ratio ${\kappa}_{\alpha\bar\beta}(s_i)/{V_X}\ll 1$.
Nevertheless, for the sake of generality we will keep $\lambda$ as a free parameter for the time being.
In our derivation of the Kahler potential we so far neglected possible higher order contributions to the 
visible sector matter Kahler potential of the form\footnote{Here we set $m_{pl}=1$ and treat all matter fields as dimensionless 
in units of $m_{pl}$.}
\be\label{deltak}
\delta \tilde K=c_{\alpha\bar\beta}(s_i)Q^{\alpha}_{c}Q^{\dagger\bar\beta}_{c}\phi_{c}\bar\phi_{c}+\,.\,.\,.=
c_{\alpha\bar\beta}(s_i)\frac{Q^{\alpha}Q^{\dagger\bar\beta}}{V_X}\frac{\phi\bar\phi}{V_X}+\,.\,.\,.\,,
\ee
which gravitationally couple the hidden sector meson to the visible sector fields $Q_{\alpha}$. In the above expression, 
the subscript $c$ denotes canonically normalized matter fields in the 4-dimensional Einstein frame.
Such couplings can create problems if the meson $F$-term is quite large (which is true in the $G_2$-MSSM)
because they can induce flavor changing neutral currents. This is the flavor problem of gravity mediated
susy breaking models\footnote{Note that the Kahler metric derived from (\ref{mkpt}) introduces no flavor problems
since, as we will see from explicit computations, the mass matrix for the scalars will be proportional to 
$\kappa_{\alpha\bar\beta}(s_i)$ and therefore diagonalization and 
canonical normalization of the kinetic terms automatically results in universal scalar masses.}. 
These terms were neglected in our previous work \cite{Acharya:2008zi}. 

Technically, computing the unknown coefficients 
$c_{\alpha\bar\beta}(s_i)$ from the underlying theory is difficult, goes well 
beyond the scope of this work and our
aim here is {\it not} to explain the flavor
structure of the supersymmetry breaking Lagrangian. Rather, we would like to understand the effect that
the presence of such terms might have on other sectors of the theory, e.g. their effect on superpartner
masses and couplings. For these purposes it is sufficient to assume that the flavor structure of the Kahler
metric is completely determined by the matrix $\kappa_{\alpha\bar\beta}(s_i)$, so that
\be\label{cdef}
c_{\alpha\bar\beta}(s_i)=\kappa_{\alpha\bar\beta}(s_i)\,\frac {c(s_i)}3\,,
\ee
where we introduced the factor of $1/3$ for future convenience. 
As we will see in the later sections, whilst this does not introduce any flavor violation, the point will be that
the effect of such terms on the mass spectrum will be similar even if we introduced
flavor violating terms, as should become clear eventually.\footnote{Generically, the absence of flavor
changing neutral currents implies that the off-diagonal entries in the mass matrix for the canonically normalized
squarks and sleptons are suppressed, though in particular models even stronger constraints are necessary, 
e.g. the requirement that the diagonal entries are nearly degenerate, depending on the spectrum and 
the $A$-terms \cite{ArkaniHamed:1997ab}.}

Actually, such a form might arise from an expansion of the Kahler potential if the 
visible and hidden sectors were completely sequestered. Though we do not expect $M$ theory to
be sequestered, it can be useful to think of the sequestering as an extreme limit in a more general
model.
 
A sequestered Kahler potential has the form
\be
K^{seq}=-3\ln\left(4\pi^{1/3}V_X-\frac 13{\phi\bar\phi}-\frac 13\kappa_{\alpha\bar\beta}Q^{\alpha}Q^{\dagger\bar\beta}\right)\,,
\ee
and the Kahler metric for the visible sector is given by
\be
K_{\alpha\bar\beta}^{seq}=\frac{\kappa_{\alpha\bar\beta}}{4\pi^{1/3}V_X-\frac 13\phi\bar\phi}\,.
\ee
Absorbing the factor of $4\pi^{1/3}$ into the definition of the fields and expanding the above expression
in powers of $\phi^2/V_X$ we obtain
\be
K_{\alpha\bar\beta}^{seq}=\frac {\kappa_{\alpha\bar\beta}}{V_X}\left(1+\frac{\phi\bar\phi}{3V_X}\right)+\,.\,.\,.\,.
\ee
Comparing the above expression with (\ref{deltak}) we can read off the coefficients
\be
c_{\alpha\bar\beta}^{seq}=\frac 13{\kappa_{\alpha\bar\beta}}\,,
\ee
which corresponds to (\ref{cdef}) when $c(s_i)=1$. Hence, function $c(s_i)$ in (\ref{cdef}) is the measure of 
deviation of the matter Kahler potential from the exactly sequestered form. As was pointed out in
\cite{Anisimov:2002az}, sequestering is not at all generic in string/$M$ theory and presumably
$G_2$ compactifications of $M$ theory are no exception. We thus will regard the value of $c(s_i)$ 
in a given vacuum as a parameter and consider the theory for various values of $c(s_i)$.

Combining all of the previous considerations, the visible sector matter Kahler metric and its inverse take the
form
\ba\label{lbl4}
&&\tilde K_{\alpha\bar\beta}=\frac {\kappa_{\alpha\bar\beta}(s_i)}{V_X}\left(1+c(s_i)\frac{\phi\bar\phi}{3V_X}\right)\,,\\
&&\tilde K^{\alpha\bar\beta}= {\kappa^{\alpha\bar\beta}(s_i)}\frac {V_X}{\left(1+c(s_i)\frac{\phi\bar\phi}{3V_X}\right)}\approx{\kappa^{\alpha\bar\beta}(s_i)} {V_X}{\left(1-c(s_i)\frac{\phi\bar\phi}{3V_X}\right)}\,,\nonumber
\ea
where $\kappa^{\alpha\bar\beta}(s_i)$ satisfies
\be
\kappa^{\alpha\bar\beta}(s_i)\kappa_{\bar\beta\gamma}(s_i)=\delta^{\alpha}_{\gamma}\,.
\ee
Combining (\ref{conp}) with the above we conclude that $\kappa^{\alpha\bar\beta}(s_i)$ is a homogeneous function of the moduli of degree $-\lambda$.
Function $c(s_i)$ will be typically assumed to take values in the range
\be
0\leq c(s_i)\leq 1\,.
\ee
However, as long as the Kahler metric is positive-definite, one may also consider the regime when $c(s_i)<0$. 
Diagonalizing the Kahler metric of the visible sector we obtain
\be
{\tilde K}_{\alpha}\delta_{\alpha\bar\beta}={\cal U}^{\dagger}_{\alpha\gamma}{\tilde K}_{\gamma\bar\rho}{\cal U}_{\bar\rho\bar\beta}\,,
\ee
where the eigenvalues ${\tilde K}_{\alpha}$ are given by
\be\label{visibleKahler}
\tilde K_{\alpha}=\frac {\kappa_{\alpha}(s_i)}{V_X}\left(1+c(s_i)\frac{\phi\bar\phi}{3V_X}\right)\,,
\ee
and $\kappa_{\alpha}(s_i)$ are homogeneous functions of degree $\lambda$ that satisfy (\ref{conp}).
In computing the anomaly mediated contribution to the gaugino masses, it will be necessary 
to compute various derivatives of $\ln\tilde K_{\alpha}$.
For this purpose, it turns out that it is very convenient to express $\ln\tilde K_{\alpha}$ as
\be\label{Ka}
\ln\tilde K_{\alpha}=\ln\kappa_{\alpha}(s_i)-\ln V_X+\ln\left(1+c(s_i)\frac{\phi\bar\phi}{3V_X}\right)\approx\frac 13 K+\ln\kappa_{\alpha}(s_i)+\left(c(s_i)-1\right)\frac{\phi\bar\phi}{3V_X}+{\rm const}\,,
\ee
where $K$ is the Kahler potential in (\ref{Kahler}).

\section{Moduli stabilization}\label{s3}

In this section we reconsider the problem of moduli stabilization with the much more general
moduli and matter Kahler potentials introduced in the previous section.
We will be working in the framework of ${\cal N}=1$ $D=4$ effective supergravity and will demonstrate that
all the moduli can be stabilized self-consistently in the regime where the supergravity approximation
is valid. Recall that in the compactifications we study here, non-Abelian gauge fields arise from 
co-dimension four singularities \cite{bsa}, \cite{Acharya:2000gb}, \cite{Atiyah:2000zz}, 
\cite{Acharya:2001hq}, \cite{Atiyah:2001qf}. In other words, there exist three-dimensional submanifolds $\cal Q$ 
inside the $G_2$-manifold $X$, along which there is an orbifold singularity of A-D-E type. 

The basic idea is that strong dynamics in the hidden sector breaks supersymmetry, stabilizes the moduli and generates a small scale.
In this context we would like to highlight some important properties that distinguish $G_2$ compactifications
from other known corners of the string landscape.  
First, unlike four-dimensional Calabi-Yau compactifications, where one typically has to deal
with several different types of moduli, e.g. complex structure, Kahler moduli, the dilaton, vector bundle moduli, etc.,
which are typically stabilized via different mechanisms, in $G_2$ compactifications of $M$ theory all deformations of the internal metric of the manifold $X$ are completely captured by the periods $s_i$ of the associative three-form $\Phi$.
Since all $s_i$ are on an equal footing the task of moduli stabilization is dramatically simplified as one can use a {\it single mechanism} to stabilize {\it all} geometric moduli\footnote{For those $G_2$ compactifications of $M$ theory that are dual to the four-dimensional vacua of the Heterotic string, the dilaton and the vector bundle moduli on the Heterotic side are mapped to some of the geometric moduli $s_i$ on the $M$ theory side.}.
Second, all the complexified moduli $z_i=t_i+is_i$ enjoy a Peccei-Quinn-type shift symmetry, which is
inherited from the gauge symmetry associated with the three-form $C_3$ of the eleven-dimensional supergravity.
In the absence of tree-level flux contributions this symmetry is exact at the perturbative level but it can be broken by non-perturbative effects. Therefore, in the fluxless sector of the theory, the entire superpotential is purely non-perturbative
and depends upon all the moduli $s_i$. Therefore, one naturally expects exponential hierarchies to be generated, 
once the moduli are stabilized\footnote{To contrast this, recall that  in Type IIB orientifold compactifications, because the complex structure moduli {\it do not} possess a shift symmetry, the superpotential generically receives unsuppressed perturbative contributions. Furthermore, with the exception of some toroidal examples, the precise dependence of the non-perturbative contributions in Type IIB orientifolds on the complex structure moduli is currently unknown. }.

The simplest possibility consistent with the supergravity approximation is a hidden sector with two gauge groups $SU(P+N_f)$ and $SU(Q)$ where the first 
is super QCD with $N_f=1$ flavor of quarks $Q$ and $\tilde Q$ transforming in 
a complex (conjugate) representation of $SU(P+1)$ (the corresponding associative cycle $\cal Q$ contains
two isolated singularities of co-dimension seven) and the second hidden sector with the gauge group $SU(Q)$
is a ``pure glue'' super Yang-Mills theory. One can easily consider more general gauge groups without
much qualitative difference.
One can also consider a setup with charged matter in both hidden
sectors. However, as was demonstrated in \cite{Acharya:2007rc}, in such cases, one of the two
$F$-terms coming from the matter fields in the hidden sectors is always suppressed relative to the other and thus does not
contribute to the quantities relevant for phenomenology. A single hidden sector gauge theory is also enough
to stabilise the moduli, though the vacuum is not in a place where supergravity is trustable!

Therefore, the non-perturbative effective 
superpotential generated by the strong gauge dynamics in the hidden sectors is given by
\be\label{supot}
W=A_1\phi^ae^{ib_1f}+A_2e^{ib_2f}\,.
\ee
The matter field $\phi$ represents an effective meson degree of freedom defined in 
(\ref{mesondef}) in terms of the chiral matter fields $\hat Q$ and $\tilde {\hat Q}$.
The coefficients $b_1$, $b_2$ and $a$ are
\be
b_1=\frac{2\pi}{P}\,,\,\,\,\,b_1=\frac{2\pi}{Q}\,,\,\,\,\,a=-\frac{2}{P}\,.
\ee

In \cite{Acharya:2007rc} it was explained that if one uses
a superpotential of the form (\ref{supot}), de Sitter vacua arise only when $Q>P$ (if we
include matter in both hidden sectors dS vacua exist without such condition).
Hence, we will keep this in mind from now on.

In (\ref{supot}) we explicitly assumed that the associative cycles supporting both hidden sectors
are in proportional homology classes which results in the gauge kinetic function being given by essentially 
the same integer combination
of the moduli $z_i$ for both hidden sectors
\be\label{gaugefunction}
f=\sum_{i=1}^{N}N_iz_i\,,
\ee
were 
\be
{\rm Im}(f)=V_{\cal Q}\equiv\int_{\cal Q}\Phi=\sum_{i=1}^{N}N_is_i
\ee
 is the volume of the corresponding associative cycle with the integers $N_i$ specifying the homology class. This possibility
may naturally arise when the three-cycle $\cal Q$ has a non-trivial fundamental group, e.g. ${\cal Q}=S^3/{\bf Z_q}$, 
so it can support discrete Wilson lines.
Then, just like the visible sector GUT is broken to the Standard Model, the unified hidden sector gauge group can be broken to a product subgroup
$SU(N+M)\rightarrow SU(N)\times SU(M)\times U(1)$ while $N+M$ and $\overline{N+M}$ matter multiplets, 
localized at two distinct co-dimension seven singularities, give rise to $(N,1)+(1,M)$ plus the conjugate\footnote{Alternatively, one may also consider a hidden $SO(2(N+M))\rightarrow SU(N)\times SU(M)\times U(1)\times U(1)$ with charged chiral matter in $2(N+M)$ giving rise to $(N,1)+(\bar N,1)+(1,M)+(1,\bar M)$.}. Unless the singularities
are extremely close, the supersymmetric mass terms of the vector-like pairs are exponentially
suppressed by the corresponding membrane instanton. Thus, one obtains two hidden supersymmetric 
QCD gauge theories with light vector-like matter supported
along the same three-cycle $\cal Q$. As mentioned above, since one of the two matter F-terms is always
suppressed relative to the other \cite{Acharya:2007rc}, one obtains virtually the same results in the
simplified scenario where one of the hidden sectors is a "pure glue" supersymmetic Yang-Mills gauge theory.

While one can certainly consider possibilities where the gauge kinetic functions $f_1$ and $f_2$ are not proportional, 
the results in \cite{Acharya:2007rc} taught us that unless $f_1\propto f_2$ it is more difficult to stabilize
all the moduli in the regime where the supergravity approximation is valid. Thus, obtaining solutions
which we can trust is the main reason for choosing to consider the case where $f_1=f_2=f$. Obviously, progress
in the more general cases would be welcome.

Typical examples for three-cycles supporting non-Abelian gauge fields in $G_2$-manifolds are spheres and their
quotients such as Lens spaces $S^3/{\bf Z_q}$ considered in \cite{Friedmann:2002ty}.
The expression in (\ref{supot}) can in principle contain many additional non-perturbative contributions
if $X$ contains other rigid associative cycles.
In that respect, the two terms included in (\ref{supot}) should be regarded as the leading order exponentials. 
As long as $Q$ and $P$ are large enough compared to the Casimirs from the other gauge groups, 
the remaining terms will be exponentially suppressed in general. This is particularly true for the membrane 
instanton corrections to (\ref{supot}) which come with exponentials containing $b_i=2\pi$.
On the other hand, some such instantons induce Yukawa interactions among the visible 
sector matter fields and are therefore implicitly assumed to be part of the full superpotential.

The total Kahler potential - moduli plus hidden sector matter, is given by
\be\label{Kahler}
K = -3 \ln 4\pi^{1/3} V_X+\kappa(s_i)\frac{\bar\phi\phi}{V_X}\,.
\ee
In what follows we first consider a simplified case where the function $\kappa(s_i)$ is a pure constant, i.e.
\be
\kappa(s_i)=1\,.
\ee
However, in section \ref{generalization} we will generalize our results to the case where $\kappa(s_i)$ is a homogeneous
function satisfying (\ref{hpr2}). The important point is that even then the functional form of the soft breaking terms
remains virtually unchanged compared to the simplified case, thus validating our approach.
In general, (\ref{Kahler}) must include the contributions to the Kahler 
potential from all matter sectors including the visible sector as described in the previous section.
However, since the visible sector fields will obtain zero vacuum expectation values (vevs), 
they can be dropped for the purposes of stabilizing moduli.

The standard ${\cal N}=1$ $D=4$ supergravity scalar potential is given by
\be
V=e^K\left(K^{n{\bar{m}}}F_n{\bar F}_{{\bar{m}}}-3|W|^2\right)\,,
\ee
where the $F$-terms are
\ba\label{Fs}
F_i&=&\partial_iW+W\partial_iK=iN_ie^{ib_2\vec N\cdot\vec t}\left(-b_1A_1\phi_0^ae^{-b_1\vec N\cdot\vec s}+b_2A_2e^{-b_2\vec N\cdot\vec s}\right)\\
&+& i\frac{3{a}_i}{2s_i}\left(1+\frac{\phi^2}{3V_X}\right)e^{ib_2\vec N\cdot\vec t}\left(-A_1\phi^ae^{-b_1\vec N\cdot\vec s}+A_2e^{-b_2\vec N\cdot\vec s}\right)\nonumber\\
F_{\phi}&=&\partial_{\phi}W+W\partial_{\phi}K=-e^{ib_2\vec N\cdot\vec t-i\theta}aA_1\phi_0^{a-1}e^{-b_1\vec N\cdot\vec s}\\\nonumber
&+& \frac{\phi_0}{V_X}e^{ib_2\vec N\cdot\vec t-i\theta}\left(-A_1\phi^ae^{-b_1\vec N\cdot\vec s}+A_2e^{-b_2\vec N\cdot\vec s}\right)\nonumber\,.
\ea
In the above we used 
\be\label{partK}
\frac{\partial K}{\partial z_i}=\frac 1{2i}\frac{\partial K}{\partial s_i}\,,
\ee
together with the definition of ${a}_i$ in (\ref{eq3}) in combination with 
\be\label{partV}
\frac{\partial }{\partial s_i}\frac 1{V_X}=\frac {\hat K_i}{3V_X}\,.
\ee
We also parameterized the meson field $\phi$ as 
\be
\phi=\phi_0e^{i\theta}\,,
\ee
and fixed one combination of the axions and the meson phase $\theta$
\be
\cos((b_1-b_2)\vec N\cdot\vec t+a\theta)=-1\,.
\ee

Before we proceed to constructing de Sitter vacua it is instructive to take a step back and consider a simpler case where the first non-perturbative term in the 
superpotential is also a pure gaugino condensate arising from a "pure glue" supersymmetric Yang-Mills theory.
In this case one possible solution corresponds to a supersymmetric AdS extremum described by the following set of equations
\ba
F_i=0\,,\,\,\,\,\Rightarrow\,\,\,\,N_i\left(-b_1A_1e^{-b_1\vec N\cdot\vec s}+b_2A_2e^{-b_2\vec N\cdot\vec s}\right)+ \frac{3{a}_i}{2s_i}\left(-A_1e^{-b_1\vec N\cdot\vec s}+A_2e^{-b_2\vec N\cdot\vec s}\right)=0\,,
\ea
which is equivalent to
\be\label{t45}
s_i=- \frac{3{a}_i}{2N_i}\frac{\left(-A_1e^{-b_1\vec N\cdot\vec s}+A_2e^{-b_2\vec N\cdot\vec s}\right)}{\left( -b_1A_1e^{-b_1\vec N\cdot\vec s}+b_2A_2e^{-b_2\vec N\cdot\vec s}\right)}\,.
\ee
Using the contraction property (\ref{eq4}) we can find from (\ref{t45}) that the volume $\vec N\cdot\vec s$ of the hidden sector three-cycle can be determined
by solving the following transcendental equation
\be
\vec N\cdot\vec s=- \frac{7}{2}\frac{\left(-A_1e^{-b_1\vec N\cdot\vec s}+A_2e^{-b_2\vec N\cdot\vec s}\right)}{\left( -b_1A_1e^{-b_1\vec N\cdot\vec s}+b_2A_2e^{-b_2\vec N\cdot\vec s}\right)}\,.
\ee
In the limit when $\vec N\cdot\vec s\gg 1$, the approximate solution is given by
\be\label{vq}
V_{\cal Q}=\vec N\cdot\vec s\approx \frac 1{b_1-b_2}\ln\left(\frac{A_1b_1}{A_2b_2}\right)>0\,,\,\,\,\,{\rm when}\,\,\,b_1>b_2\,\&\,{A_1b_1}>{A_2b_2}\,,\,\,\,{\rm or}\,\,\,b_1<b_2\,\&\,{A_1b_1}<{A_2b_2}\,.
\ee
The moduli vevs can then be found from
\be\label{ans7}
s_i=\frac{3{a}_i}{7N_i}V_{\cal Q}\,,\,\,\,\,\Rightarrow\,\,\,\,\tau_i=N_i \frac{7V_X}{3V_{\cal Q}}\,,
\ee
where the seven-dimensional volume is stabilized at
\be
V_X=V_{\cal Q}^{7/3}\left(\frac 37\right)^{7/3}V_X(s_k)\Big |_{_{s_k=\frac{{a}_k}{N_k}}}\,.
\ee
Note from (\ref{ans7}) that at the extremum, the periods of the co-associative four-form $\tau_i\sim N_i$ up to a positive constant. Recall that 
here, $N_i$ are the integers representing the homology class of  $\cal Q$:
\be
N_i=\int_{[{\cal Q}]}\phi_i\,,\,\,\,\,\,\phi_i\in H^3(X,\,Z)\,,
\ee
where the harmonic three-form $\phi_i$ is Poincare dual to a four-cycle $[\tau_i]$.
Therefore, from (\ref{ans7}) we see that extremization of the supergravity scalar potential dynamically stabilizes the
co-associative four-form  $*\Phi$ to be proportional to the integral homology class of the
associative three-cycle $\cal Q$:
\be
F_i=0\,\,\,\,\Rightarrow\,\,\,\,*\Phi=\alpha\cdot PD_{X}({\cal Q})\,,\,\,\,0<\alpha\in R\,.
\ee
Therefore, in the basis specified by (\ref{basischoice})
the integers $N_i$ must be positive definite
\be
N_i>0\,,\,\,\,\forall i=1,\,...\,\,N\,.
\ee

In order to determine the values of $a_i$ at the minimum we substitute our expressions for $s_i$ (\ref{ans7}) into the definition of ${a}_i$ in (\ref{eq3})
to get a system of $N$ transcendental equations, which then completely determine ${a}_i$ in principle
\be\label{eqforatilde}
\hat K_i\Big |_{_{s_i=\frac{{a}_i}{N_i}}}+3N_i\,=0\,.
\ee
Note that the dependence on $V_{\cal Q}$ in (\ref{eqforatilde}) is gone due to the scaling property of the volume $V_X$.
Hence, we have recast the problem of determining the moduli vevs at the minimum into a problem of determining the values of ${a}_i$.
Obviously, obtaining general analytic solutions for ${a}_i$ 
from (\ref{eqforatilde}) is impossible in practice, since $V_X$ has not been specified.
However, precisely because the moduli vevs at the minimum are given by (\ref{ans7}),
it turns out that in order to compute the quantities relevant for particle
physics, one does not need to know the values of ${a}_i$ explicitly. 
All one actually needs to know are the contraction properties (\ref{eq4}) and (\ref{contrD}).

{\it Therefore, the results we derive will be valid for any singular manifold of $G_2$ holonomy containing an associative three-cycle $\cal Q$ that
contributes to the non-perturbative superpotential in a form of at least two gaugino condensates, whose integral homology class in the basis 
(\ref{basischoice}) is specified by positive integers.} 
By explicitly checking in explicit toy examples, both numerically
and analytically it seems that, for a given form of $V_X$, an isolated solution indeed exists.

In principle, there exists an alternative way of determinimg $a_i$ more directly, although in the long run it may be more practical
to solve the system (\ref{eqforatilde}). Namely, suppose one can reexpress the volume $V_X$ in terms of the dual variables $\tau_i$ 
defined in (\ref{eq14}).\footnote{One needs to ensure that in the new basis the signature of the Hessian 
$\frac{\partial^2 V_X}{\partial \tau_i\partial \tau_j}$ remains Lorentzian.} With respect to $\tau_i$ the volume $V_X$ is a homogeneous function of degree $7/4$.
Then, we find
\be\label{eq78}
\frac{\partial V_X}{\partial \tau_i}=\sum_{j=1}^N\frac{\partial V_X}{\partial s_j}\frac{\partial s_j}{\partial \tau_i}
=\sum_{j=1}^N\frac{\partial V_X}{\partial s_j}\frac{\partial s_i}{\partial \tau_j}=\sum_{j=1}^N\tau_j\frac{\partial s_i}{\partial \tau_j}=\frac 34 s_i\,,
\ee
where we used the property that $s_i$ are homogeneous functions of $\tau_i$ of degree $3/4$ and the symmetry of the Jacobi matrix
\be
\frac{\partial \tau_i}{\partial s_j}=\frac{\partial^2 V_X}{\partial s_j \partial s_i}=\frac{\partial \tau_j}{\partial s_i}.
\ee
Then, using (\ref{eq3}), (\ref{eq78}) and (\ref{ans7}) we obtain
\be
a_i=\frac {s_i\tau_i}{V_X}=\frac {4\tau_i}{3V_X}\frac{\partial V_X}{\partial \tau_i}=\frac 43 N_i\frac{\partial \ln V_X}{\partial\tau_i}\Big |_{\tau_i=N_i}\,,
\ee
where in the final step we used the property that the "angular" variables $a_i$ do not scale. Then we can re-express
the moduli vevs (\ref{ans7}) as
\be\label{mvevs}
s_i=\frac{4}{7}V_{\cal Q}\frac{\partial \ln V_X}{\partial\tau_i}\Big |_{\tau_i=N_i}\,.
\ee

Recall that all the integers $N_i$ that describe the homology of the hidden sector associative cycle ${\cal Q}$
are fixed for a given manifold $X$. Therefore, according to (\ref{mvevs}) once we specify the microscopic details such as $V_X$ and $N_i$, the vevs of {\it all} the moduli $s_i$ are automatically determined in terms of a {\it single parameter} - the volume $V_{\cal Q}$ of the three-cycle $\cal Q$.
Therefore, all masses and couplings, being functions of $s_i$, are also fixed in terms of $V_{\cal Q}$, including  $\alpha^{-1}_{GUT}=\sum N_i^{vis}s_i=const\times V_{\cal Q}.$\footnote{In our discussion we are neglecting all the
subleading effects, e.g. the threshold corrections to $\alpha_{GUT}$ due to the Kaluza-Klein modes \cite{Friedmann:2002ty} as well as possible Coleman-Weinberg-type loop corrections to (\ref{mvevs}).}

In an explicit realistic compactification, one could automatically determine the proportionality constant between $\alpha_{GUT}^{-1}$ and $V_{\cal Q}$  from the integers $N_i^{vis}$ specifying the homology of the visible sector GUT three-cycle. Then, using the bottom-up MSSM value $\alpha_{GUT}^{-1}\approx 25$ one would be able to fix the volume $V_{\cal Q}\approx 25/const$ as well as {\it all} remaining couplings, including the visible sector Yukawa couplings and masses! Thus, given a realistic
$G_2$ compactification one could {\it in principle} make genuine predictions and quickly rule out models that do not satisfy experimental constraints\footnote{The current lack of explicit realistic $G_2$ examples presents a great challenge
in implementing these ideas. However, we shall demonstrate here that one can nevertheless make significant progress in computing
several quantities relevant for particle physics, e.g. soft terms in the supersymmetry breaking lagrangian, without 
relying on either a specific functional form of $V_X$ or the microscopic details of the MSSM embedding.}.
This extreme rigidity of fluxless $G_2$ vacua is quite remarkable and runs in stark contrast to the flexibility found for flux compactifications, where for a given manifold one can perform a very large scan over the integer fluxes
and generate distributions of masses and couplings \cite{Douglas:2003um}.

In order to illustrate how the system (\ref{eqforatilde}) is realized in practice we give a couple of explicit examples, though we
stress that we have checked many more general examples than just those given here.
Let us first consider a particularly simple $N$-parameter family of Kahler potentials consistent with $G_2$ 
holonomy where the volume $V_X$ is given by
\be\label{simplevolume}
V_X=\prod_{i=1}^N s_i^{n_i}\,,\,\,{\rm where}\,\,\,\sum_{i=1}^Nn_i=\frac 73\,.
\ee
In this case the solutions to (\ref{eqforatilde}) are simply constants given by
\be
{a}_i=n_i\,.
\ee
In fact, this example represents the class of Kahler potentials considered in the previous work \cite{Acharya:2007rc} and the
solutions are discussed in detail there.

One may consider more complicated examples such as
\be
V_X=\sum_k V_k\,,\,\,{\rm where}\,\,\,V_k\equiv c_k\prod_{i=1}^Ns_i^{n_i^k}\,,\,\,{\rm such \,\,that}\,\forall k\,\,\,\sum_{i=1}^Nn_i^k=\frac 73\,.
\ee
In this case system (\ref{eqforatilde}) translates into
\be\label{ati}
{\sum_{k}\left(n_i^k-{a}_i\right)c_k\prod_{j=1}^N\left(\frac{{a}_j}{N_j}\right)^{n_j^k}}=0\,.
\ee

In these cases one can check numerically that, for very generic sets of 
parameters $\{n_i^k,\,c_k,\,N_i\}$, the system of equations (\ref{ati}) yields positive solutions 
for ${a}_i$, where the Hessian matrix $H(V_X)_{ij}$ has Lorentzian signature\footnote{This is necessary if the homogeneous function $V_X$
is the volume of a genuine $G_2$-manifold. This also guarantees positive kinetic terms for the moduli fields.}. 
For example, choosing $N_1=1$, $N_2=1$, $N_3=1$, $N_4=1$ we numerically compute $a_i$ 
for the following toy examples with four moduli
\ba
&&V_X=s_1^{\frac 79}s_2^{\frac 79}s_3^{\frac 7{18}}s_4^{\frac 7{18}}-\frac 13s_1^1s_2^{\frac 23}s_3^{\frac 13}s_4^{\frac 13}
-\frac 12s_1^{\frac 13}s_2^1s_3^{\frac 12}s_4^{\frac 12}\,\,\Rightarrow\,\,{a}_1\approx 1.038\,,\,{a}_2\approx 0.648\,,\,{a}_3\approx 0.324\,,\,{a}_4\approx 0.324\,,\nonumber\\
&&V_X=s_2^{\frac {14}9}s_3^{\frac 7{18}}s_4^{\frac 7{18}}+\frac 13 s_1^1s_2^{\frac 23}s_4^{\frac 23}
+\frac 12 s_1^{\frac 13}s_2^1s_3^1\,\,\Rightarrow\,\,{a}_1\approx 0.051\,,\,{a}_2\approx 1.478\,,\,{a}_3\approx 0.459\,,\,{a}_3\approx 0.344\,,
\ea
where for both examples 
\be
{\rm sign}\left(\frac{\partial^2 V_X}{\partial s_i\partial s_j}\right)\Big |_{s_i=\frac{a_i}{N_i}}=(+,\,-,\,-,\,-)\,,
\ee
which explicitly demonstrates that having positive solutions for 
$a_i$ is fairly generic and more importantly is guaranteed when $V_X$ is not just a randomly picked homogeneous function of degree 
$7/3$ but represents an actual volume of a $G_2$ manifold $X$.

We now go on to consider de Sitter vacua by including the charged chiral matter fields $Q$ and $\tilde Q$ into the hidden sector.
The superpotential and the Kahler potential are given by (\ref{supot}) and (\ref{Kahler}).
In order to compute the scalar potential we need to compute the inverse
Kahler metric. Using the Kahler potential (\ref{Kahler}) together with (\ref{eq3}), (\ref{eq9}), (\ref{partK}) and 
(\ref{partV}) we first obtain the following components for the Kahler metric
\ba\label{km}
&&K_{i{\bar{j\,\,\,}}}=\frac{3{a}_{{\bar{j\,\,\,}}}}{4s_is_{{\bar{j\,\,\,}}}}\left(1+\frac{\phi_0^2}{3V_X}\right)\Delta_{i{\bar{j\,\,\,}}}
+\frac{{a}_i{a}_{{\bar{j\,\,\,}}}}{4s_is_{{\bar{j\,\,\,}}}}\frac{\phi_0^2}{V_X}\,,\\
&&K_{i\bar\phi}=i\frac{{a}_i}{2s_i}\frac{\phi}{V_X}\,,\,\,\,\,K_{\phi{\bar{j\,\,\,}}}=-i\frac{{a}_{{\bar{j\,\,\,}}}}{2s_{{\bar{j\,\,\,}}}}\frac{\bar\phi}{V_X}\,,
\,\,\,\,K_{\phi\bar\phi}=\frac 1{V_X}\,.\nonumber
\ea
Note that on the right hand side of the above expressions ${a}_{{\bar{j\,\,}}}$ and $\Delta_{i{\bar{j\,\,}}}$ 
are the same real quantities defined previously with index $j$ replaced by $\bar j$.

The inverse Kahler metric must satisfy the following set of equations
\ba
&&K^{i{\bar{j\,\,\,}}}K_{{{\bar{j\,\,}}}k}+K^{i\bar\phi}K_{\bar\phi k}=\delta^i_k\,,\\
&&K^{i{\bar{j\,\,\,}}}K_{{{\bar{j\,\,}}}\phi}+K^{i\bar\phi}K_{\bar\phi\phi}=0\,,\nonumber\\
&&K^{\phi{\bar{j\,\,\,}}}K_{{{\bar{j\,\,}}}\phi}+K^{\phi\bar\phi}K_{\bar\phi\phi}=1\nonumber\,.
\ea
After a little bit of work we obtain the following components for the inverse
Kahler metric 
\ba\label{invKahler}
&&K^{i{\bar{j\,\,\,}}}=\frac{4s_is_{\bar{j\,\,}}}{3{a}_i}\frac{(\Delta^{-1})^{i{\bar{j\,\,}}}}{1+\frac{\phi_0^2}{3V_X}}\,,
\,\,\,\,K^{i\bar\phi}=i\frac{2}{3}\frac{s_i\bar\phi}{1+\frac{\phi_0^2}{3V_X}}\,,\,\,\,\,K^{\phi{\bar{j\,\,\,}}}=-i\frac{2}{3}\frac{s_{\bar j}\phi}{1+\frac{\phi_0^2}{3V_X}}\,,\\
&&K^{\phi\bar\phi}=V_X\left(1+\frac 73\frac 1{1+\frac{\phi_0^2}{3V_X}}\frac{\phi_0^2}{3V_X}\right)\nonumber\,.
\ea
Note that despite the fact that the matter part of the Kahler potential in (\ref{Kahler}) is only given up to the quadratic order
in $\frac{\phi_0^2}{V_X}$, we decided to keep all the higher order terms inside the inverse Kahler metric.
This is self-consistent as long as the combination $\frac{\phi_0^2}{3V_X}$ appearing in the inverse Kahler 
metric is stabilized at a value sufficiently smaller than one such that the quartic and higher order terms 
are suppressed.

Now, putting all the pieces together we obtain the scalar potential
\ba\label{scpot}
V&=&\frac{e^{\phi_0^2/{V_X}}}{64\pi V_X^3}\left[\frac 43\sum_{i=1}^N\sum_{{\bar{j\,\,\,}}=1}^N\frac{s_is_{\bar{j\,\,\,}}
N_iN_{\bar{j\,\,\,}}}{a_i}\frac {(\Delta^{-1})^{i{\bar{j\,\,\,}}}}{1+\frac{\phi_0^2}{3V_X}}
\left(b_1A_1\phi_0^ae^{-b_1\vec N\cdot\vec s}
-b_2A_2e^{-b_2\vec N\cdot\vec s}\right)^2\right .\\
&+&\left .4\vec N\cdot\vec s\left(b_1A_1\phi_0^ae^{-b_1\vec N\cdot\vec s}-b_2A_2e^{-b_2\vec N\cdot\vec s}\right)
\left(A_1\phi_0^ae^{-b_1\vec N\cdot\vec s}-A_2e^{-b_2\vec N\cdot\vec s}\right)\right .\nonumber\\
&+&\left .7\left(A_1\phi_0^ae^{-b_1\vec N\cdot\vec s}-A_2e^{-b_2\vec N\cdot\vec s}\right)^2\left(1+\frac{\phi_0^2}{3V_X}\right)\right .\nonumber\\
&-&\left . \frac 43\left(\frac{b_1A_1\phi_0^ae^{-b_1\vec N\cdot\vec s}-b_2A_2e^{-b_2\vec N\cdot\vec s}}{1+\frac{\phi_0^2}{3V_X}}\vec N\cdot\vec s+\frac 72\left(A_1\phi_0^ae^{-b_1\vec N\cdot\vec s}-A_2e^{-b_2\vec N\cdot\vec s}\right)\right)\right .\nonumber\\
&\times&\left .\left(aA_1\phi_0^ae^{-b_1\vec N\cdot\vec s}+\frac{\phi_0^2}{V_X}\left(A_1\phi_0^ae^{-b_1\vec N\cdot\vec s}-A_2e^{-b_2\vec N\cdot\vec s}\right)\right)+\frac{V_X}{\phi_0^2}\left(1+\frac 73\frac 1{1+\frac{\phi_0^2}{3V_X}}\frac{\phi_0^2}{3V_X}\right)\right .\nonumber\\
&\times&\left .\left(aA_1\phi_0^ae^{-b_1\vec N\cdot\vec s}+\frac{\phi_0^2}{V_X}\left(A_1\phi_0^ae^{-b_1\vec N\cdot\vec s}-A_2e^{-b_2\vec N\cdot\vec s}\right)\right)^2\right .\nonumber\\
&-&\left .3\left(A_1\phi_0^ae^{-b_1\vec N\cdot\vec s}-A_2e^{-b_2\vec N\cdot\vec s}\right)^2\right ]\nonumber\,.
\ea
To understand the minima of the potential we will use the techniques
developed earlier in \cite{Acharya:2007rc}. Namely, we will work in the regime when the volume of the
hidden sector associative cycle $V_{\cal Q}=\vec N\cdot\vec s$ is large and expand our solutions in the
inverse powers of this volume. This is equivalent to an expansion in the UV weak hidden sector gauge coupling.
In this long a tedious procedure we utilize the methods developed in \cite{Acharya:2007rc}, yet with some 
important modifications. 

Since we are considering the simplified case by setting $\kappa(s_i)=1$ in the Kahler potential for the effective meson field, the supersymmetry
breaking $F$-term contributions are functions of $V_X$ and $\vec N\cdot\vec s$ only and therefore the scale invariant "angular" coordinates $a_i$ will remain
the same as in the supersymmetric case. On the other hand, the "radial" coordinate parameterized by $V_{\cal Q}$ (or $V_X$) will be shifted.
Reintroducing the notation of \cite{Acharya:2007rc}
\be\label{xyz}
\alpha\equiv\frac{A_1\phi_0^a}{A_2}e^{-(b1-b2)\vec N\cdot\vec s}\,,\,\,\,\,x\equiv\alpha-1\,,\,\,\,\,y\equiv b_1\alpha-b_2\,,\,\,\,\,z
\equiv b_1^2\alpha-b_2^2\,,
\ee
we therefore make the following ansatz for the moduli vevs at the minimum
\be\label{ans}
s_i=\frac{{a}_i}{N_i}\frac xyL\,.
\ee
In this notation, the volume of the associative cycles supporting the hidden sector gauge groups is given by
\be\label{VQ}
V_{\cal Q}=\vec N\cdot\vec s=\frac xyL\sum_{i=1}^N{a}_i=\frac 73\frac xyL\,,
\ee
in which case the moduli ansatz (\ref{ans}) can be rewritten as
\be\label{ans2}
s_i=\frac{{a}_i}{N_i}\frac 37V_{\cal Q}\,.
\ee
Let us first assume that $L$ is non-zero and finite when $y\rightarrow 0$. This assumption will be verified
in this section by determining $L$ explicitly. Then, we get from (\ref{VQ}) and the
definitions above
\ba
&&{V_{\cal Q}}\rightarrow\infty \,\,\,\,\Rightarrow\,\,\,\, y\rightarrow 0\,\,\,\,\Rightarrow
\,\,\,\,\alpha=\frac{b_1}{b_2}+{\cal O}(\frac 1{V_{\cal Q}})\\
&&\Rightarrow{V_{\cal Q}}=\vec N\cdot\vec s=\frac1{b_1-b_2}\ln\left(\frac{b_1A_1\phi_0^a}{b_2A_2}\right)=
\frac 1{2\pi}\frac{PQ}{Q-P}\ln\left(\frac{QA_1\phi_0^a}{PA_2}\right)
\,.\nonumber\ea
This fixes the value of the volume $V_{\cal Q}$ of the hidden sector three-cycle.

We now go on to demonstrate that the ansatz for the moduli vevs (\ref{ans}) indeed represents the correct solution at the minimum 
of the scalar potential. In particular, we must verify our assumption that $L$ is non-zero and finite in the limit $y\rightarrow 0$ 
by determining $L$ self-consistently in this limit. Hence, we will now derive the equation for $L$ and 
demonstrate explicitly that one of the possible solutions is indeed non-zero and finite in this limit. 
After minimizing the potential with respect to the moduli $s_i$ and using the definitions (\ref{xyz}) 
we obtain the following system of equations 
\ba\label{potder}
&&\frac {\partial V}{\partial s_k}=-\frac {3{a}_k}{s_k}\left(1+\frac{\phi^2}{3V_X}\right)V+\frac{e^{\phi^2/V_X}}{64\pi V_X^3}\left[\frac 43\frac{\partial}{\partial s_k}\left(\sum_{ij}\frac{s_is_jN_iN_j(\Delta^{-1})^{ij}}{{a}_i}\right)\frac {y^2}{1+\frac{\phi^2}{3V_X}}\right .\\
&&\left .-\frac 83\sum_{ij}\frac{s_is_jN_iN_j(\Delta^{-1})^{ij}}{{a}_i}\frac{N_kzy}{1+\frac{\phi^2}{3V_X}}+\frac 43\sum_{ij}\frac{s_is_jN_iN_j(\Delta^{-1})^{ij}}{{a}_i}\frac {y^2}{\left(1+\frac{\phi^2}{3V_X}\right)^2}\frac{\phi^2}{3V_X}\frac{{a}_k}{s_k}\right .\nonumber\\
&&\left .-4N_kxy-4N_k(\vec N\cdot\vec s)y^2-4N_k(\vec N\cdot\vec s)xz-7x^2\frac{\phi^2}{3V_X}\frac{{a}_k}{s_k}\right .\nonumber\\
&&\left .-2N_k\left(\frac 23\frac y{1+\frac{\phi^2}{3V_X}}-\frac 23(\vec N\cdot\vec s)\frac{z}{1+\frac{\phi^2}{3V_X}}+\frac 23(\vec N\cdot\vec s)\frac{y}{\left(1+\frac{\phi^2}{3V_X}\right)^2}\frac{\phi^2}{3V_X}\frac{{a}_k}{s_kN_k}-\frac 73y\right)\left(a\alpha+\frac{\phi^2}{V_X}x\right)\right .\nonumber\\
&&\left .+2N_k\left(\frac 23(\vec N\cdot\vec s)\frac y{1+\frac{\phi^2}{3V_X}}+\frac 73x\right)\left(b_1a\alpha+\frac{\phi^2}{V_X}y\right)\right .\nonumber\\
&&\left .+\frac{{a}_k}{s_k}\frac{V_X}{\phi^2}\left(1+\frac 73\frac 1{1+\frac{\phi^2}{3V_X}}\frac{\phi^2}{3V_X}\right)\left(a\alpha+\frac{\phi^2}{V_X}x\right)^2-\frac 79\frac{{a}_k}{s_k}\frac 1{\left(1+\frac{\phi^2}{3V_X}\right)^2}\left(a\alpha+\frac{\phi^2}{V_X}x\right)^2\right .\nonumber\\
&&\left .-2N_k\frac {V_X}{\phi^2}\left(1+\frac 73\frac 1{1+\frac{\phi^2}{3V_X}}\frac{\phi^2}{3V_X}\right)\left(a\alpha+\frac{\phi^2}{V_X}x\right)\left(ab_1\alpha+\frac{\phi^2}{V_X}\frac{{a}_k}{s_kN_k}x+\frac{\phi^2}{V_X}y\right)\right ]A_2e^{-b_2\vec N\cdot s}=0\,,\nonumber
\ea
where in one of the intermediate steps we simplified
\ba
\frac{V_X}{\phi^2}\left(\frac 73\frac 1{\left(1+\frac{\phi^2}{3V_X}\right)^2}\left(\frac{\phi^2}{3V_X}\right)^2\frac{{a}_k}{s_k}-\frac 73\frac 1{1+\frac{\phi^2}{3V_X}}\frac{\phi^2}{3V_X}\frac{{a}_k}{s_k}\right)=-\frac 79\frac 1{\left(1+\frac{\phi^2}{3V_X}\right)^2}\frac{{a}_k}{s_k}.
\ea
Multiplying (\ref{potder}) by $\frac{s_k}{{a}_k x^2}$ and using the explicit 
expression for the potential (\ref{scpot}) in terms of the quantities (\ref{xyz}) we obtain
\ba\label{Seq}
&&-3\left(1+\frac{\phi^2}{3V_X}\right)\left [\frac 43\sum_{ij}\frac{s_is_jN_iN_j(\Delta^{-1})^{ij}}{{a}_i}\frac{y^2}{x^2}\frac 1{1+\frac{\phi^2}{3V_X}}+4(\vec N\cdot\vec s)\frac yx+7\left(1+\frac{\phi^2}{3V_X}\right)\right .\\
&&\left .-2\left(\frac 23(\vec N\cdot\vec s)\frac{y/x}{1+\frac{\phi^2}{3V_X}}+\frac 73\right)\left(\frac{a\alpha}x+\frac{\phi^2}{V_X}\right)+\frac{V_X}{\phi^2}\left(1+\frac 73\frac 1{1+\frac{\phi^2}{3V_X}}\frac{\phi^2}{3V_X}\right)\left(\frac{a\alpha}x+\frac{\phi^2}{V_X}\right)^2-3\right ]\nonumber\\
&&+\frac 43\frac{s_k}{{a}_k}\frac{\partial}{\partial s_k}\left(\sum_{ij}\frac{s_is_jN_iN_j(\Delta^{-1})^{ij}}{{a}_i}\right)\frac {1}{1+\frac{\phi^2}{3V_X}}\frac{y^2}{x^2}\nonumber\\
&&-\frac 83\sum_{ij}\frac{s_is_jN_iN_j(\Delta^{-1})^{ij}}{{a}_i}\frac{zy}{x^2}\frac 1{1+\frac{\phi^2}{3V_X}}\frac{N_ks_k}{{a}_k}+\frac 43\sum_{ij}\frac{s_is_jN_iN_j(\Delta^{-1})^{ij}}{{a}_i}\frac {y^2}{x^2}\frac 1{\left(1+\frac{\phi^2}{3V_X}\right)^2}\frac{\phi^2}{3V_X}\nonumber\\
&&-4\frac{N_k s_k}{{a}_k}\frac yx-4\frac{N_ks_k}{{a}_k}(\vec N\cdot\vec s)\frac{y^2}{x^2}-4\frac{N_ks_k}{{a}_k}(\vec N\cdot\vec s)\frac zx-7\frac{\phi^2}{3V_X}\nonumber\\
&&-2\frac{N_ks_k}{{a}_k}\left(\frac 23\frac {y/x}{1+\frac{\phi^2}{3V_X}}-\frac 23(\vec N\cdot\vec s)\frac{z/x}{1+\frac{\phi^2}{3V_X}}+\frac 23(\vec N\cdot\vec s)\frac{y/x}{\left(1+\frac{\phi^2}{3V_X}\right)^2}\frac{\phi^2}{3V_X}\frac{{a}_k}{s_kN_k}-\frac 73\frac yx\right)\left(\frac{a\alpha}x+\frac{\phi^2}{V_X}\right)\nonumber\\
&&+2\frac{N_ks_k}{{a}_k}\left(\frac 23(\vec N\cdot\vec s)\frac {y/x}{1+\frac{\phi^2}{3V_X}}+\frac 73\right)\left(\frac{b_1a\alpha}x+\frac{\phi^2}{V_X}\frac yx\right)\nonumber\\
&&+\frac{V_X}{\phi^2}\left(1+\frac 73\frac 1{1+\frac{\phi^2}{3V_X}}\frac{\phi^2}{3V_X}\right)\left(\frac{a\alpha}x+\frac{\phi^2}{V_X}\right)^2-\frac 79\frac 1{\left(1+\frac{\phi^2}{3V_X}\right)^2}\left(\frac{a\alpha}x+\frac{\phi^2}{V_X}\right)^2\nonumber\\
&&-2\frac{N_ks_k}{{a}_k}\frac {V_X}{\phi^2}\left(1+\frac 73\frac 1{1+\frac{\phi^2}{3V_X}}\frac{\phi^2}{3V_X}\right)\left(\frac{a\alpha}x+\frac{\phi^2}{V_X}\right)\left(\frac{ab_1\alpha}x+\frac{\phi^2}{V_X}\frac{{a}_k}{s_kN_k}+\frac{\phi^2}{V_X}\frac yx\right)=0\,.\nonumber
\ea
At first sight it appears that finding an analytic expression for $L$ from (\ref{Seq}) is hopeless since a closed form for
$(\Delta^{-1})^{i{\bar{j\,\,\,}}}$ is unknown and ${a}_i$ have not been determined explicitly. However, upon further
examination we notice that in order to find $L$ from (\ref{Seq}) we only need to know the contraction rules (\ref{eq4}), (\ref{eq7}) and (\ref{contrD}).
Indeed, using the the ansatz (\ref{ans}) together with the definition (\ref{eq5}) and applying (\ref{eq4}), (\ref{eq7}) and (\ref{contrD}) we first evaluate the terms
\ba
&&\frac{s_k}{{a}_k}\frac{\partial}{\partial s_k}\left(\sum_{ij}\frac{s_is_jN_iN_j(\Delta^{-1})^{ij}}{{a}_i}\right)=\frac{x^2}{y^2}L^2\frac 1{{a}_k}\sum_j(\Delta^{-1})^{kj}{a}_j+\frac{x^2}{y^2}L^2\sum_i(\Delta^{-1})^{ik}\\
&&+\frac{x^2}{y^2}L^2\frac 1{{a}_k}\sum_i\left(\frac{P_{ik}}{{a}_i}\sum_j(\Delta^{-1})^{ij}{a}_j\right)+\frac{x^2}{y^2}L^2\sum_j\left({a}_j\sum_i\frac{\partial (\Delta^{-1})^{ij}}{\partial s_k}\right)\nonumber\\
&&=2\frac{x^2}{y^2}L^2
+\frac{x^2}{y^2}L^2\frac 1{{a}_k}\sum_iP_{ik}+\frac{x^2}{y^2}L^2\sum_j\left({a}_j\frac{\partial}{\partial s_k}\sum_i(\Delta^{-1})^{ij}\right)=2\frac{x^2}{y^2}L^2\,,\nonumber\\
&&\sum_{ij}\frac{s_is_jN_iN_j(\Delta^{-1})^{ij}}{{a}_i}=\frac{x^2}{y^2}L^2\sum_i\sum_j(\Delta^{-1})^{ij}{a}_j=\frac{x^2}{y^2}L^2\sum_i{a}_i=\frac 73\frac{x^2}{y^2}L^2\,,
\ea
and then use the same ansatz (\ref{ans}) and contraction identities for the rest of the terms in (\ref{Seq}) to obtain the following equation for $L$
\ba
&&-3\left(1+\frac{\phi^2}{3V_X}\right)\left [\frac {28}9 L^2\frac 1{1+\frac{\phi^2}{3V_X}}+\frac{28}3L+7\left(1+\frac{\phi^2}{3V_X}\right)\right .\\
&&\left .-2\left(\frac {14}9\frac L{1+\frac{\phi^2}{3V_X}}+\frac 73\right)\left(\frac{a\alpha}x+\frac{\phi^2}{V_X}\right)+\frac{V_X}{\phi^2}\left(1+\frac 73\frac 1{1+\frac{\phi^2}{3V_X}}\frac{\phi^2}{3V_X}\right)\left(\frac{a\alpha}x+\frac{\phi^2}{V_X}\right)^2-3\right ]\nonumber\\
&&+\frac 83\frac {L^2}{1+\frac{\phi^2}{3V_X}}
-\frac {56}9\frac{zx}{y^2}\frac {L^3}{1+\frac{\phi^2}{3V_X}}+\frac {28}9\frac {L^2}{\left(1+\frac{\phi^2}{3V_X}\right)^2}\frac{\phi^2}{3V_X}\nonumber\\
&&-4L-\frac{28}3L^2-\frac{28}3\frac {zx}{y^2}L^2-7\frac{\phi^2}{3V_X}\nonumber\\
&&-2L\left(\frac 23\frac {1}{1+\frac{\phi^2}{3V_X}}-\frac {14}9\frac{L}{1+\frac{\phi^2}{3V_X}}\frac{zx}{y^2}+\frac {14}9\frac{L}{\left(1+\frac{\phi^2}{3V_X}\right)^2}\frac{\phi^2}{3V_X}-\frac 73\right)\left(\frac{a\alpha}x+\frac{\phi^2}{V_X}\right)\nonumber\\
&&+2L\left(\frac {14}9\frac {L}{1+\frac{\phi^2}{3V_X}}+\frac 73\right)\left(\frac{b_1a\alpha}y+\frac{\phi^2}{V_X}\right)\nonumber\\
&&+\frac{V_X}{\phi^2}\left(1+\frac 73\frac 1{1+\frac{\phi^2}{3V_X}}\frac{\phi^2}{3V_X}\right)\left(\frac{a\alpha}x+\frac{\phi^2}{V_X}\right)^2-\frac 79\frac 1{\left(1+\frac{\phi^2}{3V_X}\right)^2}\left(\frac{a\alpha}x+\frac{\phi^2}{V_X}\right)^2\nonumber\\
&&-2L\frac {V_X}{\phi^2}\left(1+\frac 73\frac 1{1+\frac{\phi^2}{3V_X}}\frac{\phi^2}{3V_X}\right)\left(\frac{a\alpha}x+\frac{\phi^2}{V_X}\right)\left(\frac{ab_1\alpha}y+\frac{\phi^2}{V_X}\right)\nonumber\\
&&-2\left(1+\frac 73\frac 1{1+\frac{\phi^2}{3V_X}}\frac{\phi^2}{3V_X}\right)\left(\frac{a\alpha}x+\frac{\phi^2}{V_X}\right)=0\,,\nonumber
\ea
Multiplying the above equation by $\frac 3{28}(1+\frac{\phi^2}{3V_X})\frac{y^2}{zx}$ and taking the limit $y\rightarrow 0$ we obtain
\ba\label{cuLeq}
&&\frac{2}3L^3+L^2\left(1-\frac{a\alpha}{3x}\right)-L^2\frac{b_1a\alpha y}{3xz}-L\frac{b_1a\alpha y}{2xz}\left(1+\frac{\phi_0^2}{3V_X}\right)+L\frac{3b_1a\alpha y}{14xz}\left(1+\frac{10}9\frac{\phi_0^2}{V_X}\right)\left(\frac{a\alpha V_X}{\phi_0^2x}+1\right)=0\,,\nonumber
\ea
where we dropped terms of ${\cal O}(y^2)$ and higher. 
A non-trivial solution can be obtained by solving the corresponding quadratic equation
\ba\label{Leq}
&&\frac{2}3L^2+L\left(1-\frac{a\alpha}{3x}\right)-L\frac{b_1a\alpha y}{3xz}-\frac{b_1a\alpha y}{2xz}\left(1+\frac{\phi_0^2}{3V_X}\right)+\frac{3b_1a\alpha y}{14xz}\left(1+\frac{10}9\frac{\phi_0^2}{V_X}\right)\left(\frac{a\alpha V_X}{\phi_0^2x}+1\right)=0\,
\ea
which is analogous to the equation in the second line in (126) of \cite{Acharya:2007rc}.

Solving (\ref{Leq}) to the first subleading order in $y$ results in
\be\label{Lsol}
L=-\frac 32\left(1-\frac{a\alpha}{3x}\right)+y\frac{3b_1a\alpha }{14xz}\frac{1+\frac{a\alpha V_X}{\phi_0^2x}}{1-\frac{a\alpha}{3x}}\left(1+\frac{\phi_0^2}{3V_X}\right)\,.
\ee
Hence, we see that this solution is non-zero and finite when $y\rightarrow 0$ and therefore is
self-consistent. This is the solution describing the minimum of the potential. 
We must note that there is another possible solution of (\ref{Leq}) for which 
$L\sim y\rightarrow 0$. In fact this other solution corresponds to the extremum at the top of the potential 
barrier and we will not discuss it further.
Using (\ref{Lsol}) we can now compute the first subleading order correction to $\alpha$ to obtain
\ba\label{alphasub}
\alpha&=&\frac PQ+\frac{7P\left(3(Q-P)-2\right)}{12\pi Q}\frac 1{V_{\cal Q}}\\
&=&\frac PQ+\frac{7(Q-P)^2}{2Q^2}\left(1-\frac 2{3(Q-P)}\right)\frac P{P_{eff}}\nonumber\,,
\ea
where we have introduced
\be\label{peffdef}
P_{eff}\equiv P\ln\left(\frac{QA_1\phi_0^a}{PA_2}\right)\,.
\ee
Using (\ref{alphasub}) we can express the solution for $L$ from (\ref{Lsol}) as
\ba
L&=&-\frac 32\left(1-\frac 2{3(Q-P)}\right)+\frac 7{2P_{eff}}\left(1-\frac 2{3(Q-P)}\right)\\
&+&\frac 3{2P_{eff}}\left(1+\frac{2V_X}{(Q-P)\phi_0^2}\right)\left(1+\frac{\phi_0^2}{3V_X}\right)\nonumber\,.
\ea
In the leading order, the moduli vevs are given by
\be\label{modvev}
s_i=\frac{{a}_i}{N_i}\frac{3QP_{eff}}{14\pi(Q-P)}\,.
\ee
We note that since ${a}_i$, $N_i$ are positive, we need $P_{eff}>0$ if $Q>P$, so that there exists a
local minimum with $s_i>0$. 

The next step is to determine the vev of the effective meson field by minimizing the
potential with respect to $\phi_0$. Let us first compute the potential at the minimum as a function
of the meson. The result is given in equation (\ref{vacen}) and the reader not interested in its derivation
may proceed directly there.
It turns out that since the moduli vevs at the minimum are proportional to ${a}_i/N_i$ as in (\ref{ans2}),
explicit computation of the $F$-terms at the minimum and various contractions thereof while using the
rules (\ref{eq4}) and (\ref{contrD}) becomes possible.
Let us demonstrate some of these computations in detail. First we need to identify the gravitino mass
in terms of our notation in (\ref{xyz}). Using the usual definition in combination with (\ref{xyz}) we have
\be
m_{3/2}=e^{K/2}|W|=e^{K/2}|x|A_2e^{-b_2\vec N\cdot\vec s}\,.
\ee
Because the existence of de Sitter vacua requires $Q-P>0$ (see \cite{Acharya:2007rc} for details)
we obtain using (\ref{alphasub}) that
\be
x\approx \frac PQ-1<0\,.
\ee
On the other hand, since $m_{3/2}>0$ we can express the following combination in terms of the gravitino mass
\be\label{eq15}
e^{K/2}xA_2e^{-b_2\vec N\cdot\vec s}=-m_{3/2}\,.
\ee
We now multiply $F_i$ in (\ref{Fs}) by $e^{K/2}$ and using (\ref{xyz}) and (\ref{eq15}) express
\ba\label{eq16}
e^{K/2}F_i&=&iN_ie^{i\gamma_W}\left(-y-x\frac{3{a}_i}{2s_iN_i}\left(1+\frac{\phi_0^2}{3V_X}\right)\right)e^{K/2}A_2e^{-b_2\vec N\cdot\vec s}\\
&=&iN_ie^{i\gamma_W}\left(\frac yx+\frac{3{a}_i}{2s_iN_i}\left(1+\frac{\phi_0^2}{3V_X}\right)\right)m_{3/2}\,,\nonumber
\ea
where $\gamma_W$ denotes the overall phase of the superpotential.
Using the ansatz (\ref{ans}) for $s_i$ we obtain from (\ref{eq16})
\ba\label{eq17}
e^{K/2}F_i&=&iN_ie^{i\gamma_W}\left(\frac yx+\frac{3y}{2xL}\left(1+\frac{\phi_0^2}{3V_X}\right)\right)m_{3/2}\,\\
&=&iN_ie^{i\gamma_W}\frac 7{3V_{\cal Q}}\left(L+\frac{3}{2}\left(1+\frac{\phi_0^2}{3V_X}\right)\right)m_{3/2}\,,\nonumber
\ea
where in the second line we used
\be\label{eq18}
\frac xyL=\frac 37 V_{\cal Q}\,,
\ee
obtained from (\ref{VQ}). Similarly, we find from (\ref{Fs}) using (\ref{xyz}) together with (\ref{eq15})
\be\label{eq19}
e^{K/2}F_{\phi}=e^{i(\gamma_W-\theta)}\left(\frac{a\alpha}{\phi_0x}+\frac{\phi_0}{V_X}\right)m_{3/2}\,.
\ee
Before computing $e^{K/2}F^i$ we would like to express the $K^{i{\bar{j\,\,}}}$ components of the inverse Kahler
metric at the minimum using the ansatz (\ref{ans}) for $s_{{\bar{j\,\,}}}$ as follows
\ba
K^{i{\bar{j\,\,\,}}}=\frac{4s_is_{\bar{j\,\,\,}}}{3{a}_i}\frac{(\Delta^{-1})^{i{\bar{j\,\,}}}}{1+\frac{\phi_0^2}{3V_X}}
=\frac {xL}y\frac{4s_i{a}_{\bar{j\,\,}}}{3{a}_iN_{\bar{j\,\,\,}}}\frac{(\Delta^{-1})^{i{\bar{j\,\,\,}}}}{1+\frac{\phi_0^2}{3V_X}}
={V_{\cal Q}}\frac{4s_i{a}_{\bar{j\,\,}}}{7{a}_iN_{\bar{j\,\,\,}}}\frac{(\Delta^{-1})^{i{\bar{j\,\,\,}}}}{1+\frac{\phi_0^2}{3V_X}}\,.
\ea
Contracting (\ref{eq17}) and (\ref{eq19}) with the inverse Kahler metric and using the solution for $L$ from (\ref{Lsol}) we then obtain
\ba\label{eq20}
e^{K/2}F^i&=&e^{K/2}K^{i{\bar{j\,\,}}}\bar F_{{\bar{j\,\,}}}+e^{K/2}K^{i\bar\phi}\bar F_{\bar\phi}\\
&=&-ie^{-i\gamma_W}\frac {4 s_i}{3{a}_i}\sum_{\bar j=1}^N{a}_{\bar{j\,\,\,}}(\Delta^{-1})^{i{\bar{j\,\,\,}}}\left(\frac L{1+\frac{\phi_0^2}{3V_X}}+\frac 32\right)m_{3/2}
+ie^{-i\gamma_W}\frac 23\frac{s_i}{1+\frac{\phi_0^2}{3V_X}}\left(\frac{a\alpha}x+\frac{\phi_0^2}{V_X}\right)m_{3/2}\nonumber\\
&=&-is_ie^{-i\gamma_W}\frac{2yb_1a\alpha }{7xz}\frac{1+\frac{a\alpha V_X}{\phi_0^2x}}{1-\frac{a\alpha}{3x}}m_{3/2}\approx-ie^{-i\gamma_W}\frac{2s_i}{P_{eff}}\left({1+\frac{a\alpha V_X}{\phi_0^2x}}\right)m_{3/2}\,,\nonumber
\ea
where in the last line we used (\ref{alphasub}) to plug into $x$, $y$, and $z$ defined by (\ref{xyz}) except for the 
combination $\left({1+\frac{a\alpha V_X}{\phi_0^2x}}\right)$ and kept the leading term in $1/P_{eff}$.
Note that in order to get from the second to third line
in (\ref{Fup}) we used the second contraction property in (\ref{contrD}).
Similarly, contracting (\ref{eq17}) and (\ref{eq19}) with the corresponding components of the inverse
Kahler metric (\ref{invKahler}) we obtain
\be\label{eq21}
e^{K/2}F^{\phi}=e^{-i\gamma_W}\phi\left(1-\frac 7{3P_{eff}}\right)\left({1+\frac{a\alpha V_X}{\phi_0^2x}}\right)m_{3/2}\,.
\ee

Using the results (\ref{eq17}), (\ref{eq19}), (\ref{eq20}) and (\ref{eq21}) together with (\ref{Lsol}) and (\ref{alphasub})
we can compute the following contributions
\ba\label{eq22}
&&e^KF^iF_i=\frac 7{P_{eff}}\left(\frac{a\alpha}x+\frac{\phi_0^2}{V_X}\right)^2
\left(\frac{V_X}{\phi_0^2}\right)^2\left[\frac{\phi_0^2}{3V_X}
+\frac 1{P_{eff}}\left(1+\frac{\phi_0^2}{3V_X}\right)\right]m_{3/2}^2\\
&&e^KF^{\phi}F_{\phi}=\left(\frac{a\alpha}x+\frac{\phi_0^2}{V_X}\right)^2\frac{V_X}{\phi_0^2}\left(1-\frac 7{3P_{eff}}\right)m_{3/2}^2\,,\nonumber
\ea
where we also used $\vec N\cdot\vec s=V_{\cal Q}$ while performing the computations in the first line of (\ref{eq22}).

Then, the potential at the minimum is given by
\ba
V_0&=&e^K(F^iF_i+F^{\phi}F_{\phi}-3|W|^2)\\
&=&\left(\frac{a\alpha}x+\frac{\phi_0^2}{V_X}\right)^2\frac{V_X}{\phi_0^2}m_{3/2}^2+\frac 7{P_{eff}^{\,2}}\left(\frac{a\alpha}x+\frac{\phi_0^2}{V_X}\right)^2\left(1+\frac{\phi_0^2}{3V_X}\right)\left(\frac{V_X}{\phi_0^2}\right)^2m_{3/2}^2-3m_{3/2}^2\,.\nonumber
\ea
Using (\ref{alphasub}) and dropping the terms of order ${\cal O}(1/P_{eff}^2)$ we obtain the following expression for the leading contribution to the vacuum energy as a function of the meson field
\be\label{vacen}
V_0=\left[\left(\frac 2{Q-P}+\frac{\phi_0^2}{V_X}\right)^2+\frac {14}{P_{eff}}\left(1-\frac 2{3(Q-P)}\right)\left(\frac 2{Q-P}+\frac{\phi_0^2}{V_X}\right)-3\frac{\phi_0^2}{V_X}\right]\frac{V_X}{\phi_0^2}m_{3/2}^2\,.
\ee
The polynomial in the square brackets in (\ref{vacen}) is quadratic with respect to the canonically normalized meson vev squared $\phi_{c}^2\equiv\phi_0^2/V_X$ with the coefficient of the $(\phi_0^2/V_X)^2$ monomial being positive (+1) and therefore, the minimum $V_0$ is positive when the corresponding discriminant is negative. Tuning the cosmological constant to zero is then equivalent to setting the discriminant of the above polynomial to zero, which boils down to a simple condition
\be\label{peffanalyt}
P_{eff}=\frac{14(3(Q-P)-2)}{3(3(Q-P)-2\sqrt{6(Q-P)})}\,.
\ee 
Note that $P_{eff}$ defined in (\ref{peffdef}) is actually dependent on $\phi$ but because of the 
smallness of $a$ and the Log dependence, it was safe to use the approximation $P_{eff}\approx const$. 
This approximation turned out to be self-consistent since $P_{eff}$ is fairly large.
From (\ref{peffanalyt}) we see immediately that
\be
P_{eff}>0\,\Rightarrow\,Q-P\geq 3\,.
\ee
Minimizing (\ref{vacen}) with respect to $\phi_{c}^2$ and imposing the condition that the expression in the square 
brackets in (\ref{vacen}) is tuned to zero, we obtain the meson vev at the minimum in the leading order
\be
\phi_{c}^2=\frac{\phi_0^2}{V_X}\approx\frac 2{Q-P}+\frac 7{P_{eff}}\left(1-\frac 2{3(Q-P)}\right)\,.
\ee
If we tune the leading contribution to the vacuum energy and set $Q-P=3$ we obtain
\be
P_{eff}\approx 63.5\,,\,\,\,\,\frac{\phi_0^2}{V_X}\approx 0.75\,.
\ee
Recalling the factor of two in the definition of the meson field (\ref{mesondef}) we
note that along the D-flat direction, the bilinears of the canonically normalized charged matter fields that appear in 
the original Kahler potential have a somewhat smaller vev
\be<Q_c{ Q}^{\dagger}_c>=<{{\tilde Q}}_c{ {\tilde Q}}^{\dagger}_c>
=< Q_c{ {\tilde Q}}_c>\approx 0.37\,m_{pl}^2\,,
\ee
which makes it a bit easier to justify the truncation of the higher order terms in the Kahler potential 
for hidden sector matter.

We find numerically that for the minimum value $Q-P=3$, the tuning of the cosmological 
constant by varying the constants $A_1$ and $A_2$ inside the superpotential
results in fixing the value of $P_{eff}$ at
\be\label{CCcond}
P_{eff}\approx 61.648\,,
\ee
while the canonically normalized meson vev squared is stabilized at
\be\label{mvev}
\phi_{c}^2=\frac{\phi_0^2}{V_X}\approx 0.746\,,
\ee
thus confirming the analytical results above. For example, we obtain the values in (\ref{CCcond})
and (\ref{mvev}) by minimizing the scalar potential numerically for the following toy example with two moduli
\ba
&&P=27\,,\,\,Q=30\,,\,\,A_1=27\,,\,\,A_2=2.1544\,,\,\,N_1=N_2=1\,,V_X=s_1^{\frac 76}s_2^{\frac 76}
+\frac 13s_1^1s_2^{\frac 43}+\frac 12s_1^{\frac 13}s_2^2\nonumber\\\nonumber\\
&&\,\,\,\,\,\,\,\,\,\,\,\,\,\,\,\,\,\,\,\,\,\,\,\,\,\,\,\,\,\,\,\,\,\,\,\,\,\,\,\,\,\,\,\,\,\,\,\,\,\,\,\,\,\,\,\,\,\,\,\,\Rightarrow s_1\approx 34.52\,,\,\,\,s_2\approx 63.13\,.\nonumber
\ea

It is instructive to compare the moduli vevs obtained above numerically with the values obtained
by using the analytic expression (\ref{modvev}). However, before we can apply (\ref{modvev})
we need to determine the values of ${a}_i$ at the minimum. This can be done
by plugging the values of $N_i$, $n_i^k$ and $c_k$ into the system (\ref{ati}) and solving it
numerically. To compute $s_i$ from (\ref{modvev}) we use $P_{eff}$ from (\ref{CCcond}) in order
to get a better accuracy. As a result, we obtain: 
\ba
&& a_1\approx 0.825\,,\,\,\,a_2\approx 1.51\,,\,\,\,s_1\approx 34.7\,,\,\,\,s_2\approx 63.4\,,\nonumber
\ea
which confirms explicitly that the analytic expression (\ref{modvev}) for the moduli vevs at the minimum is indeed very accurate and reliable. Here we also
verified that the Hessian of the volume has Lorentzian signature.

Although the value in (\ref{mvev}) is not much smaller than one, the combination $\frac{\phi_0^2}{3V_X}$
inside the inverse Kahler metric (\ref{invKahler}) has a value
\be
\frac{\phi_0^2}{3V_X}\approx 0.25\,,
\ee
which is small enough to make the quartic and higher order terms which we kept inside the 
inverse Kahler metric much smaller. 

As we will see in the computations that follow, the value of $P_{eff}$ will enter into many 
quantities relevant for particle physics, such as tree-level gaugino masses, etc. 
Here we note that small changes in $P_{eff}$ do not affect the supersymmetry breaking masses
much, but do change the  cosmological constant significantly.  
For instance, while changing the value of 
$P_{eff}$ in the range $61\leq P_{eff}\leq 62$ hardly affects the values of the soft breaking terms,
as will be evident from the corresponding explicit formulas, such small changes in $P_{eff}$ result in 
vastly different values of the vacuum energy:
\be\label{ed45}
61\leq P_{eff}\leq 62\,\,\,\,\Rightarrow\,\,\,\,-\left(m_{3/2}m_{pl}\right)^2\times 10^{-3}\lesssim V_0\lesssim+\left(m_{3/2}m_{pl}\right)^2\times 10^{-3}\,. 
\ee
Therefore, once we coarsely tune $P_{eff}$ to its approximate value, the cosmological constant problem becomes
completely decoupled from the rest of particle physics. Even though this should be the case, it is satisfying
to see it explicitly in a complete example of moduli coupled to matter.

Note also that in the original paper \cite{Acharya:2007rc} we obtained $P_{eff}\approx 83$. This is
due to the different matter Kahler potential considered there.
As we will see, this numerical difference will result in slightly different values for the soft 
breaking terms if compared to those obtained in \cite{Acharya:2007rc},\cite{Acharya:2008zi}.

Recall that for a stable minimum to exist it is necessary that $Q-P\geq 3$. We have seen that
when $Q-P=3$ and the minimum of the potential is approximately tuned to zero, the value of $P_{eff}\approx 60$
which ensures that the moduli (\ref{modvev}) can be reliably fixed at values large enough to satisfy the
supergravity approximation. On the other hand, when $Q-P=4$, from (\ref{peffanalyt}) we get $P_{eff}\approx 20$,
in which case if $Q$ is fixed, the moduli vevs become smaller by about a factor of four. Thus,
unless the ranks of hidden sector gauge groups are incredibly large, situations when $Q-P>3$
may put our solutions well outside of the supergravity approximation. Therefore, from now on we will only
consider the case when $Q-P=3$ to ensure the validity of the regime where 
our construction is reliable.

\section{Masses and soft supersymmetry breaking terms}
\subsection{Gravitino mass}
In supergravity the bare gravitino mass is defined as
\be
m_{3/2}=m_{pl}e^{K/2}|W|=e^{K/2}|x|A_2e^{-b_2V_{\cal Q}}\,,
\ee
and can now be computed since we stabilized $V_{\cal Q}$ explicitly. It is given by
\be\label{grmass}
m_{3/2}=m_{pl}\frac{e^{\frac{\phi_0^2}{2V_X}}}{8\sqrt{\pi}V_X^{3/2}}|P-Q|\frac{A_2}Q
e^{-\frac{P_{eff}}{Q-P}}\,.
\ee

When the cosmological constant is tuned such that (\ref{CCcond}) is satisfied, for $Q-P=3$ we obtain
\be
m_{3/2}\approx 9\times 10^5({\rm TeV})\frac{C_2}{V_X^{3/2}}\,,
\ee
where $C_2\equiv A_2/Q$ was defined in (\ref{constants}). 
Calculating $C_2$ goes beyond the scope of this paper. Here we will treat $C_2$ as a phenomenological parameter
with values $C_2\sim{\cal O}(0.1-1)$ since it may experience a mild exponential suppression as in
(\ref{constants}).

On the other hand, the actual value at which the volume $V_X$ must be stabilized can be almost uniquely
determined from the scale of Grand Unification. In particular, we can use equation (4.12) in \cite{Friedmann:2002ty}
to express
\be
G_N=\frac 1{8\pi m_p^2}=\frac{\alpha_{GUT}^3V_Q^{7/3}L({\cal Q})^{2/3}}{32\pi^2M_{GUT}^2V_X}\,,
\ee
where the factor $L({\cal Q})$ is due to the threshold corrections 
from the Kaluza-Klein modes and is given by
\be
L({\cal Q})=4q\sin^2\left(5\pi w/q\right)\,,
\ee
such that $5w$ is not divisible by $q$. For typical values
\be
\alpha_{GUT}=\frac 1{V_{\cal Q}}=\frac 1{25}\,,\,\,\,M_{GUT}= 2\times 10^{16}\,{\rm GeV}\,,
\ee
we obtain 
\be
V_X=137.4\times L({\cal Q})^{2/3}\,.
\ee

In Table I we list a few typical benchmark values for the volume
and the resulting gravitino mass up to the overall factor $C_2$.
\begin{table}[t!]
\label{volumeandmgr}
\begin{tabular}{||c||c|c|c|c|c|c|c||}
\hline\hline  \rule{0pt}{3.0ex}\rule[-1.5ex]{0pt}{0pt} &
Point~1 & Point~2 & Point~3 & Point~4 & Point~5 & Point~6  & Point~7  \\ \hline
$q$ \rule{0pt}{3.0ex} & $\qquad$ 2 $\qquad$ & $\qquad$ 3 $\qquad$  & $\qquad$ 4 $\qquad$  & $\qquad$ 4 $\qquad$  & $\qquad$ 6 
$\qquad$ & $\qquad$ 6 $\qquad$ & $\qquad$ 6 $\qquad$\\
$w$ \rule{0pt}{3.0ex} & 1 & 1,\,2 & 1,\,3 & 2 & 1,\,5 & 2,\,4 & 3\\
\hline
$V_X$ & 549.6 & 594.5 & 549.6 & 872.4 & 453.7 & 943.7 & 1143.2\\
${m_{3/2}}/{C_2}$ & 70\,{\rm TeV} & 62\,{\rm TeV} &  70\,{\rm TeV} 
&  35\,{\rm TeV} &  93.0\,{\rm TeV} & 31\,{\rm TeV} &  23\,{\rm TeV}\\
\hline\hline
\end{tabular}%
\caption{Typical values of $V_X$ and ${m_{3/2}}$ divided by $C_2$ for different values of $q$ and $w$.}
\end{table}

Interestingly, the gravitino mass scale naturally turns out to be
constrained to $m_{3/2}\sim {\cal O}(10){\rm TeV}$. While this is presumably large enough
to alleviate the gravitino problem, it is also small enough to give {\it some} of the superpartners
masses which can be easily accessible at the LHC energies. As we will see below,
this is possible because of the significant suppression of the tree-level gaugino masses relative to 
$m_{3/2}$.

In addition to the gravitino mass it is instructive to compute the scale of gaugino condensation.
Using (\ref{CCcond}) the volume of the hidden sector associative cycle for $Q-P=3$ is given by
\be
V_{\cal Q}=\frac{QP_{eff}}{2\pi(Q-P)}\approx\frac{10Q}{\pi}\,,
\ee
From (\ref{StrongScale}) the scale of gaugino condensation in the second hidden sector is
\be
\Lambda\sim m_{pl}\frac{e^{-\frac{2\pi}{3Q}V_{\cal Q}}}{2{\pi}^{1/6}V_{X}^{1/2}}
\approx m_{pl}\frac{e^{-20/3}}{2{\pi}^{1/6}V_X^{1/2}}\approx \frac{1.1\times 10^{14}{\rm GeV}}{L({\cal Q})^{1/3}}\,.
\ee

\subsection{Moduli masses}
In order to compute the masses of the moduli we first need to evaluate the matrix $V_{mn}$
with $m,n=\overline{1,N+1}$, with components given by
\be
V_{ij}=\frac{\partial^2V}{\partial s_i\partial s_j}\,,\,\,\,\,
V_{i\,N+1}=\frac{\partial^2V}{\partial s_i\partial\phi_0}\,,\,\,\,\,
V_{N+1\,N+1}=\frac{\partial^2V}{\partial\phi_0\partial\phi_0}\,,
\ee
at the minimum of the potential. However, because the Kahler metric in (\ref{km}) is not diagonal,
we also need to find a unitary transformation $U$ which diagonalizes the Kahler metric. We denote
all the components of the Kahler metric as $K_{m\bar n}$. Then, by diagonalizing $K_{m\bar n}$ we obtain
\be
K_k\delta_{k\bar l}=U^{\dagger}_{k m}K_{m\bar n}U_{\bar n\bar l}\,.
\ee
After that, we need to rescale the fluctuations of the moduli around the minimum by the corresponding
$1/\sqrt{2K_k}$ factors so that the new {\it real} scalar fields have canonical kinetic terms.
At the end, finding the moduli mass squared eigenvalues boils down to diagonalizing the following matrix
\be
M_{kl}^2=\frac 12\frac1{\sqrt{K_kK_l}}U^{\dagger}_{k m}V_{mn}U_{nl}\,.
\ee
Unlike most of the other masses, the detailed form of the moduli mass matrix does depend upon the
detailed form of $V_X$. Therefore we have resorted to numerical analyses in this case
and found that there is one 
heavy modulus whose mass mainly depends on $Q$ and for $Q=30$
\be
M\sim O(200-300)\times m_{3/2}\,,
\ee
and $N$ lighter moduli with masses
\be\label{light}
m_i\sim O(1)\times m_{3/2}\,,\,\,\,\,i=\overline{1,N}\,.
\ee
The heavy modulus arises from the fluctuation which deforms the volume of the three-cycle $V_{\cal Q}$,
while $N$-1 light moduli originate from the fluctuations approximately preserving the volume and tangential
to the hyperplane defined by
\be
\vec N\cdot \vec s-V_{\cal Q}=0\,.
\ee
The remaining light modulus represents the fluctuations
of the hidden sector meson $\phi$ mixed with the geometric moduli.

\subsection{Gaugino masses}
The universal tree-level contribution to the gaugino masses can be computed from the standard supergravity
formula \cite{Brignole:1997dp}
\be
m_{1/2}^{tree}=\frac{e^{K/2}F^i\partial_if_{vis}}{2i{\rm Im}f_{vis}}\,,
\ee
where the visible sector gauge kinetic function is another integer combination
of the moduli
\be\label{gaugekineticvis}
f_{vis}=\sum_{i=1}^NN_i^{vis}z_i\,.
\ee

Note that, since the dominant $F$-term is that of the meson field, the gaugino masses at tree level
will be suppressed wrt the gravitino mass. Since the scalar masses typically get contributions of
order $m_{3/2}$ the expectation is to have light gauginos and heavier scalars, as we will indeed verify shortly.

Plugging the solution for $\alpha$ (\ref{alphasub}) into (\ref{eq20}) while using the definitions
(\ref{xyz}) we obtain
\ba\label{Fup}
e^{K/2}F^i\approx -i\frac{2s_i}{P_{eff}}\left(1+\frac{2V_X}{(Q-P)\phi_0^2}\right)m_{3/2}\,,
\ea
where we dropped the overall phase factor $e^{-i\gamma_W}$.
It is now straightforward to compute the tree-level gaugino mass
\be\label{gm}
m_{1/2}^{tree}\approx -\frac{1}{P_{eff}}\left(1+\frac{2V_X}{(Q-P)\phi_0^2}+{\cal O}\left(\frac 1{P_{eff}}\right)\right)m_{3/2}\,.
\ee
It is interesting to note that this 
formula is identical to the leading order
expression previously obtained in \cite{Acharya:2007rc} when one replaces the
combination $\phi_0^2/V_X$ by the canonically normalized meson field. Here, again the suppression coefficient
is completely independent of the number of moduli $N$ as well as the integers $N_i$ ($N_i^{vis}$) appearing inside
either the hidden sector (\ref{gaugefunction}) or the visible sector (\ref{gaugekineticvis})
gauge kinetic functions. Moreover, all the detailed dependence on the individual
moduli is completely buried inside the volume $V_X$ and the gravitino 
mass $m_{3/2}$ (which also depends on $V_X$) and therefore expression 
(\ref{gm}) is universally valid for any $G_2$ manifold that yields positive solutions of the system
of equations in (\ref{eqforatilde}).
Hence, despite the presence of a huge number of unknown microscopic parameters, the tree-level
gaugino masses in (\ref{gm}) depend on very few of them.
Moreover, when the cosmological constant is tuned to a small value and $Q-P=3$, the gaugino mass
suppression coefficient becomes completely fixed! Indeed, using (\ref{CCcond}) and 
(\ref{mvev}) for $Q-P=3$ we obtain
\be\label{gmtree1}
m_{1/2}^{tree}\approx -0.0307\times m_{3/2}\,.
\ee
This result gets slightly corrected by the threshold corrections
to the gauge kinetic function from the Kaluza-Klein modes
computed in \cite{Friedmann:2002ty}
\be
\alpha_{GUT}^{-1}=f_{vis}+\frac 5{2\pi}{\cal T}_{\omega}\,.
\ee
In the above formula, ${\cal T}_{\omega}$ is a topological invariant (Ray-Singer torsion)
\be\label{rstorsion}
{\cal T}_{\omega}=\ln\left(4\sin^2(5\pi w/q)\right)\,,
\ee
where $w$ and $q$ are integers such that $5w$ is not divisible by $q$.
In this case, the tree-level gaugino mass is given by
\be\label{gmtree}
m_{1/2}^{tree}\approx -0.0307\,\eta\times m_{3/2}\,,
\ee
where
\be\label{eta}
\eta=1-\frac {5\,g_{GUT}^2}{8\pi^2}{\cal T}_{\omega}\,.
\ee
\subsection{Anomaly mediated contribution to the gaugino masses}
Because of the substantial suppression of the universal tree-level gaugino mass, it makes
sense to take into account the anomaly mediated contributions which appear at one-loop.
The anomaly mediated contributions are given by the following general expression \cite{Bagger:1999rd}
\be\label{amgeneral}
m_a^{AM}=-\frac{g_a^2}{16\pi^2}\left[-\left(3C_a-\sum_{\alpha} C_a^{\,\alpha}\right)e^{K/2}{\overline W}+
\left(C_a-\sum_{\alpha} C_a^{\alpha}\right)e^{K/2}F^nK_n+2\sum_{\alpha}\left(C^{\,\alpha}_a
e^{K/2}F^n\partial_n\ln\tilde K_{\alpha}\right)\right]\,,
\ee
where $C_a$ and $\sum_{\alpha}C_a^{\,\alpha}$ are the quadratic Casimirs of the a-th gauge group
and $\tilde K_{\alpha}$ are eigenvalues of the Kahler metric for the visible sector fields (\ref{visibleKahler}).
Assuming the MSSM particle content, we have the following values for the Casimirs
\ba\label{casimirs}
&&U(1):\,\,\,\,\,\,\,\,C_a=0\,\,\,\,\,\,\,\,\,\,\,\,\sum_{\alpha} C_a^{\,\alpha}=\frac{33}{5}\\
\nonumber\\
&&SU(2):\,\,\,\,C_a=2\,\,\,\,\,\,\,\,\,\,\,\,\sum_{\alpha} C_a^{\,\alpha}=7\nonumber\\
\nonumber\\
&&SU(3):\,\,\,\,C_a=3\,\,\,\,\,\,\,\,\,\,\,\,\sum_{\alpha} C_a^{\,\alpha}=6\,.\nonumber
\ea

Plugging the solution for $\alpha$ (\ref{alphasub}) into (\ref{eq21}) while using the definitions
(\ref{xyz}) we obtain
\be\label{FphiUp}
e^{K/2}F^{\phi}\approx \phi\left(1-\frac 7{3P_{eff}}\right)\left(1+\frac{2V_X}{(Q-P)\phi_0^2}\right)m_{3/2}\,,
\ee
where we dropped the overall phase factor $e^{-i\gamma_W}$.
Combining (\ref{Fup}), (\ref{FphiUp}) and using (\ref{Kahler}), (\ref{visibleKahler}), (\ref{Ka}) and (\ref{conp}) we 
now compute the contributions
\ba\label{FK}
&&e^{K/2}F^nK_n=e^{K/2}F^iK_i+e^{K/2}F^{\phi}K_{\phi}=\left(\frac{\phi_0^2}{V_X}+\frac 7{P_{eff}}\right)
\left(1+\frac{2V_X}{(Q-P)\phi_0^2}\right)m_{3/2}\\
&&e^{K/2}F^n\partial_n\ln\tilde K_{\alpha}=\frac 13\left(1+\frac{2V_X}{(Q-P)\phi_0^2}\right)\left(c(s_i)\frac{\phi_0^2}{V_X}
+\frac{ 7-3\lambda}{P_{eff}}\right)m_{3/2}\,.\nonumber
\ea
In the above we also used (\ref{partK}) and (\ref{partV}) together with the definition of ${a}_i$ 
in (\ref{eq3}) as well as its contraction property (\ref{eq4}). We also dropped unknown subleading contributions
proportional to $e^{K/2}F^i\partial_i c(s_i)\sim m_{1/2}^{tree}s_i\partial_i c(s_i)$.

Using the definition (\ref{amgeneral}) we then obtain the following expression for the anomaly mediated contributions to the gaugino masses
\ba
m_a^{AM}&\approx&\frac{\alpha_{GUT}}{4\pi}\left [\left(3C_a-\sum_{\alpha} C_a^{\,\alpha}\right)
\left(1-\frac 13\left(1+\frac{2V_X}{(Q-P)\phi_0^2}\right)\left(\frac{\phi_0^2}{V_X}+\frac 7{P_{eff}}\right)\right)\right .\\
&+&\left .\frac 23\left(1+\frac{2V_X}{(Q-P)\phi_0^2}\right)\left[\left(1-c(s_i)\right)\frac{\phi_0^2}{V_X}+\frac{3\lambda}{P_{eff}}\right]
\sum_{\alpha}C^{\,\alpha}_a\right ]\times m_{3/2}\nonumber\,,
\ea
where we have explicitly separated the conformal anomaly contribution from the Konishi anomaly term using (\ref{Ka}). 

Notice the appearance of the function $c(s_i)$ which controls the size of the higher order corrections to the
matter Kahler potential. As expected, when $\lambda=0$ the Konishi anomaly vanishes in the exactly sequestered case
\cite{Choi:2007ka}, i.e. when $c(s_i)=1$. 
Again, in the leading order in $1/P_{eff}$, when $c(s_i)=0$ the result obtained above is almost the same as
the one in \cite{Acharya:2007rc}. 
Just like in the case with tree-level gaugino masses, the above result is completely
independent of the detailed moduli dependence of the volume $V_X$ and therefore
is completely general. In what follows, we will regard the value of the function $c(s_i)$
at the minimum of the scalar potential as a phenomenological parameter
\be\label{fp}
c\equiv c(s_i)\,.
\ee

When we set $\lambda=0$ and $Q-P=3$, tune the leading contribution to the vacuum energy by imposing the constraint (\ref{CCcond}), 
use (\ref{mvev}) and combine the above formula with the tree-level contribution (\ref{gmtree}), 
we obtain the following expression for the total gaugino masses
\be
M_{a}\approx \left[-0.0307\,\eta+\alpha_{GUT}\left(0.0364\left(3C_a-\sum_{\alpha} C_a^{\,\alpha}\right)
+0.0749\left(1-c\right)\sum_{\alpha} C_a^{\,\alpha}\right)\right]\times m_{3/2}\,.
\ee
Note that as was previously pointed out in \cite{Choi:2007ka}, in the limit when $c\rightarrow 1$ we obtain a 
particular type of a mirage pattern for gaugino masses \cite{Choi:2005ge}. 
However, as we will see below, in this limit the scalars become tachyonic and therefore, the 
exact mirage pattern is disfavored.
An exact numerical computation confirms the above result giving
\be\label{mamed}
M_{a}\approx \left[-0.03156\,\eta+\alpha_{GUT}\left(0.034086\left(3C_a-\sum_{\alpha} C_a^{\,\alpha}\right)
+0.07926\left(1-c\right)\sum_{\alpha} C_a^{\,\alpha}\right)\right]\times m_{3/2}\,.
\ee
Substituting the MSSM Casimirs (\ref{casimirs}) into (\ref{mamed}) we then obtain
\ba\label{gmassnew}
&&M_1\approx \left(-0.03156\,\eta+\alpha_{GUT}\left(-0.22497+0.52313\,(1-c)\right)\right)\times m_{3/2}\\
&&M_2\approx\left(-0.03156\,\eta+\alpha_{GUT}\left(-0.03409+0.55483\,(1-c)\right)\right)\times m_{3/2}\nonumber\\
&&M_3\approx\left(-0.03156\,\eta+\alpha_{GUT}\left(0.10226+0.47557\,(1-c)\right)\right)\times m_{3/2}\,.\nonumber
\ea 
The form of (\ref{gmassnew}) allows us to see explicitly that for $c=0$ the Konishi anomaly contribution is
larger than the contribution from the conformal anomaly by a factor of a few, which is what made the gaugino mass spectrum
in \cite{Acharya:2008zi} very different from other known patterns. However, as we will see below, suppressing the scalar masses 
relative to the gravitino mass by tuning the coefficient $c$
will automatically result in a large suppression of the Konishi anomaly.

\subsection{Scalars}
The masses of the unnormalized scalars can be computed from the following general expression \cite{Brignole:1997dp}
\be\label{defsc}
m^{\prime 2}_{\alpha\bar\beta}=\left(m_{3/2}^2+V_0\right){\tilde K}_{\alpha\bar\beta}-e^KF^n{\bar F}^{\bar m}\left(\partial_{\bar m}\partial_n{\tilde K}_{\alpha\bar\beta}
-\partial_{\bar m}{\tilde K}_{\alpha\bar\gamma}{\tilde K}^{\bar\gamma\delta}\partial_n{\tilde K}_{\delta\bar\beta}\right)\,.
\ee
Since the vacuum energy is tuned to zero we set $V_0=0$ in the above. Using (\ref{lbl4}),  (\ref{Kahler}),  (\ref{partK}),  (\ref{partV}), the contraction
properties (\ref{eq1}),  (\ref{eq2}),  (\ref{conp})  and the F-terms (\ref{Fup}), (\ref{FphiUp}) we obtain from (\ref{defsc}) in the leading order
\be\label{scm}
m^{\prime 2}_{\alpha\bar\beta}\approx(1-c(s_i))\left(m_{3/2}^2-\frac 73 \left(m_{1/2}^{tree}\right)^2\right){\tilde K}_{\alpha\bar\beta}+\lambda\left(m_{1/2}^{tree}\right)^2{\tilde K}_{\alpha\bar\beta}\,,
\ee
where for consistency reasons we only kept contributions linear in $c(s_i)$ and dropped unknown subleading terms proportional to 
the derivatives of $c(s_i)$, such as  e.g.
$e^KF^i{\bar F}^{\bar j}\partial_i\partial_{\bar j}c(s_i)\sim \left(m_{1/2}^{tree}\right)^2s_is_j\partial_i\partial_j c(s_i)$ . In the above derivation we also used the following properties
\be
e^{K}F^i{\bar F}^{\bar j}\partial_i\partial_{\bar j}{\hat K}=7 \left(m_{1/2}^{tree}\right)^2\,\,\,\Rightarrow\,\,\,e^{K}F^n{\bar F}^{\bar m}\partial_n\partial_{\bar m}\frac{\phi\bar\phi}{3V_X}=m_{3/2}^2-\frac 73\left(m_{1/2}^{tree}\right)^2\,.
\ee

Notice that despite the presence of the derivatives of the Kahler metric ${\tilde K}_{\alpha\bar\beta}$  in the definition (\ref{defsc}),
 the final expression (\ref{scm}) contains ${\tilde K}_{\alpha\bar\beta}$ only as an overall multiplicative factor. This happened
because the moduli F-terms (\ref{Fup}) up to a phase are essentially given by $e^{K/2}F^i= 2 s_i\times m_{1/2}^{tree}$ 
and the matrix $\kappa_{\alpha\bar\beta}(s_i)$ is a homogeneous function satisfying (\ref{conp}). Therefore, diagonalization
and canonical normalization of the corresponding kinetic terms automatically results in universal masses for
the canonically normalized scalars
\be\label{sq1}
m_{\alpha}^2\approx(1-c(s_i))\left(m_{3/2}^2-\frac 73 \left(m_{1/2}^{tree}\right)^2\right)+\lambda\left(m_{1/2}^{tree}\right)^2\,.
\ee
After setting $\lambda=0$, we tune the leading contribution to the vacuum energy by imposing the constraint (\ref{CCcond}) and 
use (\ref{gmtree1}) to obtain from (\ref{sq1})
\be
m_{\alpha}\approx(1-c)^{1/2}0.999\,m_{3/2}\,,
\ee
where we again treat the value of the function $c(s_i)$ for a given vacuum as a phenomenological parameter (\ref{fp}).
A numerical computation in this case gives excellent agreement
\be\label{scl}
m_{\alpha}\approx(1-c)^{1/2}0.998\,m_{3/2}\approx(1-c)^{1/2}\,m_{3/2}\,.
\ee
Again, for $c=0$ we recover the old result in \cite{Acharya:2007rc} where all the scalars have a flavor-universal 
mass equal to the gravitino mass.

Furthermore, the anomaly contributions to the scalar mass squareds are suppressed 
relative to the gravitino mass and since we wish to consider generic ${\cal O}(1)$ values of $(1-c)$
we will neglect such contributions.
Concretely  we are going to consider only those values of $0<c<1$ which give
\be
\frac 1{16\pi^2}<<\frac{m_{\alpha}}{m_{3/2}}\,,
\ee
such that the anomaly mediated contributions to the scalar masses can be safely neglected.
However, one can certainly extend our model and include such contributions.
Once again, the result above is completely independent of the details of $V_X$
and therefore holds for any $G_2$ manifold that solves the system (\ref{eqforatilde})
with ${a}_i>0$ such that the Kahler metric at the minimum is positive definite.

\subsection{Trilinear couplings}
The unnormalized trilinear couplings for the visible sector fields can be computed from the following general expression \cite{Brignole:1997dp}
\be\label{tri}
A^{\prime}_{\alpha\beta\gamma}=\frac{{\overline W}}{|W|}e^{K/2}F^m\left[K_mY^{\prime}_{\alpha\beta\gamma}+\partial_m Y^{\prime}_{\alpha\beta\gamma}
-\left({\tilde K}^{\delta\bar\rho}\partial_m{\tilde K}_{\bar\rho\alpha} Y^{\prime}_{\delta\beta\gamma}+(\alpha\leftrightarrow\beta)+(\alpha\leftrightarrow\gamma)\right)\right]\,,
\ee

where $\{\alpha\,,\beta\,,\gamma\}$ label visible sector matter fields and $Y^{\prime}_{\alpha\beta\gamma}$ are the 
unnormalized Yukawas that appear in the superpotential.
Recall that the Yukawa couplings $Y^{\prime}_{\alpha\beta\gamma}$ arise from the membrane instantons wrapping
associative cycles  $Q^{\alpha\beta\gamma}$, which connect isolated singularities supporting the corresponding 
matter multiplets. They are given by
\be\label{Yprime}
Y^{\prime}_{\alpha\beta\gamma}=C_{\alpha\beta\gamma}e^{i2\pi\sum_im_i^{\alpha\beta\gamma}z_i}\,.
\ee
The integer combination of the moduli $V_{Q^{\alpha\beta\gamma}}=\sum_i m_i^{\alpha\beta\gamma}s_i$ gives
the volume of the associative cycle $Q^{\alpha\beta\gamma}$ connecting co-dimension seven
singularities $\alpha$, $\beta$ and $\gamma$ where the chiral multiplets are localized. 
The coefficients $C_{\alpha\beta\gamma}$ are constants. The relation between the
physical and unnormalized Yukawa couplings is given by 
\be\label{Y}
Y_{\alpha\beta\gamma}=\frac{{\overline W}}{|W|}e^{K/2}Y^{\prime}_{\alpha\beta\gamma}\left({\tilde K}_{\alpha}{\tilde K}_{\beta}{\tilde K}_{\gamma}\right)^{-1/2}\,.
\ee
Using (\ref{Fup}) and (\ref{Yprime}) we can compute the contribution
\be\label{dY}
e^{K/2}F^m\partial_mY^{\prime}_{\alpha\beta\gamma}=Y^{\prime}_{\alpha\beta\gamma}\frac{4\pi}{P_{eff}}\left(1+\frac{2V_X}{(Q-P)\phi_0^2}\right)V_{Q^{\alpha\beta\gamma}}m_{3/2}\,.
\ee
Similarly, using (\ref{lbl4}),  (\ref{Kahler}),  (\ref{partK}),  (\ref{partV}), the contraction
properties (\ref{eq1}),  (\ref{eq2}),  (\ref{conp})  and the F-terms (\ref{Fup}), (\ref{FphiUp}) we find
\be\label{kdY}
e^{K/2}F^m{\tilde K}^{\delta\bar\rho}\partial_m{\tilde K}_{\bar\rho\alpha}=\frac {\delta^{\delta}_{\alpha}}3\left(1+\frac{2V_X}{(Q-P)\phi_0^2}\right)\left(c(s_i)\frac{\phi_0^2}{V_X}+\frac {7-3\lambda}{P_{eff}}\right)m_{3/2}\,,
\ee
where for consistency reasons we only retained contributions linear in $c(s_i)$ and dropped  unknown subleading terms
proportional to $e^{K/2}F^i\partial_i c(s_i)\sim m_{1/2}^{tree}s_i\partial_i c(s_i)$.
Again, in the above expressions we did not display the overall phase factor $e^{-i\gamma_W}$.
Using the definition (\ref{tri}) along with (\ref{FK}), (\ref{dY}) and (\ref{kdY}) we obtain the following expression for
the physical (normalized) trilinear couplings at tree-level
\be
A_{\alpha\beta\gamma}=Y_{\alpha\beta\gamma}\left(1+\frac{2V_X}{(Q-P)\phi_0^2}\right)\left((1-c(s_i))\frac{\phi_0^2}{V_X}+\frac {3\lambda+4\pi V_{Q^{\alpha\beta\gamma}}}{P_{eff}}\right)m_{3/2}\,,
\ee
which gets reduced to the result in \cite{Acharya:2007rc} when $c=0$ and $\lambda=0$. Once again, the detailed 
structure of the volume $V_X$ played absolutely no role in our ability to obtain
the above expression for the tree-level trilinear couplings. The actual volumes of three-cycles 
$V_{Q^{\alpha\beta\gamma}}$ do depend on the microscopic properties of $G_2$ manifolds
and in our general framework these parameters remain undetermined.
However, below we will present a good argument for dropping such volume
contributions completely when the third generation trilinear couplings are computed.

Setting $\lambda=0$ and $Q-P=3$, when the leading contribution to the vacuum energy is tuned we obtain for the reduced trilinears (the physical
trilinears divided by the physical Yukawa couplings)
\be\label{tricouplnew12}
{\tilde A}_{\alpha\beta\gamma}\equiv\frac{A_{\alpha\beta\gamma}}{Y_{\alpha\beta\gamma}}=\left(1.41(1-c)+0.386\times V_{Q^{\alpha\beta\gamma}}\right)m_{3/2}\,.
\ee
From the corresponding numerical calculation we obtain the following result
\be\label{tricouplnew1}
{\tilde A}_{\alpha\beta\gamma}
=\left(1.494(1-c)+0.3966\times V_{Q^{\alpha\beta\gamma}}\right)m_{3/2}\,.
\ee

Since the physical Yukawa couplings for the third generation fermions are much larger than
the first two generation Yukawas, one can typically neglect the trilinears for the first 
and second generations. Moreover, the large size of the third generation Yukawas implies
that the volumes of the three-cycles of the corresponding membrane instantons are very small.
In fact, because the top Yukawa is of order one, one can assume that the point 
$p_1$ supporting the up-type Higgs ${\bf 5}$ of $SU(5)$ coincides with the point $p_2$ supporting the third
generation ${\bf 10}$, so that the coupling $H_u{\bf 10_3}{\bf 10_3}$ has no exponential suppression 
\cite{Atiyah:2001qf}, \cite{Friedmann:2002ty}. 
At the same time the point $p_3$ supporting the down-type ${\bf \bar 5}$ Higgs and the point
$p_4$ supporting the third generation matter ${\bf \bar 5}$ are distinct but still close to $p_2\,,p_1$ so that the 
coupling of $H_d{\bf \bar 5_3}{\bf 10_3}$ which accounts for the bottom(tau) Yukawa is slightly smaller than the top Yukawa at the GUT scale.
These considerations completely justify dropping the corresponding $V_{Q^{\alpha\beta\gamma}}$ terms for the third
generation trilinears which then become simplified
\be\label{tricouplnew}
{\tilde A}_t={\tilde A}_b={\tilde A}_{\tau}\approx 1.494(1-c)\,m_{3/2}\,.
\ee
For generic values of $c$ the trilinears are of the same order as the gravitino mass.
In the limit $c\rightarrow 1$, the reduced trilinear couplings at tree-level become suppressed relative to the gravitino
mass. Note that as $c$ approaches one, the suppression of the trilinear couplings above is much stronger than that of the scalars.
In this case, the anomaly-mediated contributions may become comparable to the tree-level ones and therefore must be taken into account. General expressions given in 
\cite{Gaillard:2000fk} can be simplified in the nearly sequestered limit as
\be\label{atrigen}
{\tilde A}_a^{AM}=-\frac 1{16\pi^2}\gamma_a\left(e^{K/2}{\overline W}-\frac 13e^{K/2}F^nK_n\right)+\frac {\left(1-c\right)}{16\pi^2}X_a\,m_{3/2}\,,
\ee
where the last term denotes the unknown contributions vanishing in the sequestered limit. Note that
such terms are suppressed compared to the tree-level piece (\ref{tricouplnew}) due to the loop factor. 
As long as $(1-c)$ is small enough, they become subleading and we will drop them in further
analysis. Using (\ref{FK}) and substituting the corresponding MSSM expressions for $\gamma_a$s,
where we set $g_1=g_2=g_3=g_{GUT}$, we obtain the following expressions for the anomaly mediated
contributions to the reduced trilinear couplings
\ba\label{atri}
&&{\tilde A}_t^{AM}\approx-\frac 1{16\pi^2}\left(-\frac{46}5g^2_{GUT}+6Y_t^2+Y_b^2\right)\left(1-\frac 13\left(1+\frac{2V_X}{(Q-P)\phi_0^2}\right)\left(\frac{\phi_0^2}{V_X}+\frac 7{P_{eff}}\right)\right)m_{3/2}\,\\
&&{\tilde A}_b^{AM}\approx-\frac 1{16\pi^2}\left(-\frac{44}5g^2_{GUT}+Y_t^2+6Y_b^2+Y_{\tau}^2\right)\left(1-\frac 13\left(1+\frac{2V_X}{(Q-P)\phi_0^2}\right)\left(\frac{\phi_0^2}{V_X}+\frac 7{P_{eff}}\right)\right)m_{3/2}\,\nonumber\\
&&{\tilde A}_{\tau}^{AM}\approx-\frac 1{16\pi^2}\left(-\frac{24}5g^2_{GUT}+3Y_b^2+4Y_{\tau}^2\right)\left(1-\frac 13\left(1+\frac{2V_X}{(Q-P)\phi_0^2}\right)\left(\frac{\phi_0^2}{V_X}+\frac 7{P_{eff}}\right)\right)m_{3/2}\,.\nonumber
\ea
When we set $Q-P=3$, tune the tree-level vacuum energy by imposing the constraint (\ref{CCcond}), 
use (\ref{mvev}) and combine the above formula with the tree-level contribution (\ref{tricouplnew1}), 
we obtain
\ba\label{atri1}
&&{\tilde A}_t\approx 1.41(1-c)m_{3/2}-0.0029\left(-\frac{46}5g^2_{GUT}+6Y_t^2+Y_b^2\right)\,m_{3/2}\,\\
&&{\tilde A}_b\approx 1.41(1-c)m_{3/2}-0.0029\left(-\frac{44}5g^2_{GUT}+Y_t^2+6Y_b^2+Y_{\tau}^2\right)\,m_{3/2}\,\nonumber\\
&&{\tilde A}_{\tau}\approx  1.41(1-c)m_{3/2}-0.0029\left(-\frac{24}5g^2_{GUT}+3Y_b^2+4Y_{\tau}^2\right)\,m_{3/2}\,.\nonumber
\ea
Numerical computations give the following expressions for the total reduced trilinears
\ba\label{atri2}
&&{\tilde A}_t\approx 1.494(1-c)m_{3/2}-0.0027\left(-\frac{46}5g^2_{GUT}+6Y_t^2+Y_b^2\right)\,m_{3/2}\,\\
&&{\tilde A}_b\approx 1.494(1-c)m_{3/2}-0.0027\left(-\frac{44}5g^2_{GUT}+Y_t^2+6Y_b^2+Y_{\tau}^2\right)\,m_{3/2}\,\nonumber\\
&&{\tilde A}_{\tau}\approx  1.494(1-c)m_{3/2}-0.0027\left(-\frac{24}5g^2_{GUT}+3Y_b^2+4Y_{\tau}^2\right)\,m_{3/2}\,,\nonumber
\ea
which demonstrate a fairly high accuracy of the analytically derived result in (\ref{atri1}). 

\subsection{$\mu$ and $B\mu$ -terms}

The full hidden sector plus visible sector Kahler potential and superpotential can
be written in the following general form
\ba\label{genkahler}
&&K_{\rm total}=K\left(s_i,\phi,\bar\phi\right)+\tilde K_{\alpha\bar\beta}(s_i,\phi,\bar\phi)Q^{\alpha} Q^{\dagger\bar\beta}+Z_{\alpha\beta}(s_i,\phi,\bar\phi)Q^{\alpha}Q^{\beta}+\,h.\,c.\,\\
&&\hat W=W_{np}+\mu^{\prime}Q^{\alpha}Q^{\beta}+Y^{\prime}_{\alpha\beta\gamma}Q^{\alpha}Q^{\beta}Q^{\gamma}+\,.\,.\,.\,.\nonumber
\ea
Here, $\phi$ denote the hidden sector matter fields while $Q^{\alpha}$ are visible sector chiral matter fields where $\tilde K_{\alpha\bar\beta}(s_i,\phi,\bar\phi)$ is the visible sector Kahler metric and $Y^{\prime}_{\alpha\beta\gamma}$ are the corresponding unnormalized Yukawa couplings. 
It can be shown that the supersymmetric mass parameter $\mu^{\prime}$ can be forbidden 
by requiring certain discrete symmetries which are also used in order to solve the problem of 
doublet-triplet splitting \cite{Witten:2001bf}. 
Hence, in our analysis we will rely on the Giudice-Masiero mechanism \cite{Giudice:1988yz} in generating
effective $\mu$ and $B\mu$ terms where the bilinear coefficient $Z_{\alpha\beta}(s_i,\phi,\bar\phi)$ in (\ref{genkahler}) plays
a key role. The general expressions for the normalized $\mu$ and $B\mu$ are given by \cite{Brignole:1997dp}
\ba \label{mubmu}
\mu &=&
\left(\frac{{\overline W_{np}}}{|{W_{np}}|}e^{{K}/2}{\mu}'+m_{3/2}Z-e^{{K}/2}F^{\bar{m}}\partial_{\bar{m}}Z\right)\,
(\tilde{K}_{H_u}\tilde{K}_{H_d})^{-1/2}  \\\nonumber\\
B\mu &=&\left [\frac{{\overline W_{np}}}{|{W_{np}}|} e^{{K}/2}
{\mu}'(e^{{K}/2}F^m\,[{K}_m\,+\partial_m\ln{\mu}'-\partial_m\ln(\tilde{K}_{H_u}\tilde{K}_{H_d})]-m_{3/2})\right .\,\nonumber\\\nonumber\\
&+&\left .(2m_{3/2}^2+V_0)\,Z-m_{3/2}e^{K/2}F^{\bar m}\partial_{\bar m}Z+m_{3/2}e^{{K}/2}F^m(\partial_m Z-Z\partial_m\ln(\tilde{K}_{H_u}\tilde{K}_{H_d}))\right .\,\nonumber\\\nonumber\\
&-&\left .e^K F^{\bar m}F^n(\partial_{\bar m}\partial_n\,Z-\partial_{\bar m}Z\partial_n\ln(\tilde{K}_{H_u}\tilde{K}_{H_d}))\right ] (\tilde{K}_{H_u}\tilde{K}_{H_d})^{-1/2}\,.\nonumber
\ea
where we can set $\mu^{\prime}=0$. Unfortunately, at this point we do not have a reliable way to compute the 
Higgs bilinear $Z_{\alpha\beta}(s_i,\phi,\bar\phi)$ for $G_2$ compactifications.
Therefore, in our analysis we will parameterize the $\mu$ and $B\mu$ terms as follows
\ba\label{muterm}
\mu &=&Z^{\,1}_{eff}\,m_{3/2} \\
B\mu &=&Z^{\,2}_{eff}\,m_{3/2}^2\,,\nonumber
\ea
and treat $Z^{\,1}_{eff}$ and $Z^{\,2}_{eff}$ as phenomenological parameters.
Naturally, we expect that $Z^{\,1,\,2}_{eff}\sim{\cal O}(1)$ and, as we will see in the next section, 
tuning $\mu$ parameter in order to get the correct value of the Z - boson mass boils down
to tuning the values of $Z^{\,1,\,2}_{eff}$.

\section{Generalization to the case when $\kappa(s_i)$ is a non-trivial homogeneous function}\label{generalization}

Recall that the above results have been obtained assuming that the factor $\kappa(s_i)$ appearing in the Kahler potential 
(\ref{Kahler}) of the hidden sector matter is a pure constant. In this section we briefly outline the main results
for the case when $\kappa(s_i)$ is a general homogeneous function of the moduli of degree zero, satisfying
the property (\ref{hpr2}). Here we will not give any explicit analytic derivations (these are quite tedious and would mostly resemble
the computations in the preceding sections) and instead present numerical evidence that most of results
obtained for the simplified case $\kappa(s_i)=1$ can be directly extended to the more general scenario.

The main difference from the previous case is in the form of the moduli vevs at the minimum of the scalar potential.
These are now given by
\be\label{simod}
s_i\approx\frac 1{N_i}\left({a}_i+c_i\frac{\phi_c^2}3 r\right)\frac{3QP_{eff}}{14\pi(Q-P)}\,,
\ee
where $r\approx 3/2$ when $Q-P=3$. Note that the vev $\phi_c^2$ of the canonically normalized effective meson field at the minimum
is given by the same expression as in the case when $\kappa(s_i)=1$:
\be\label{scf}
\phi_c^2\equiv\kappa(s_i)\frac{\phi_0^2}{V_X}\approx\frac 2{Q-P}+\frac 7{P_{eff}}\left(1-\frac 2{3(Q-P)}\right)\,,
\ee
where $P_{eff}$ is exactly the same as in (\ref{peffanalyt}). Keep in mind that the analytic expression (\ref{scf}) is only 
valid when the leading contribution to the vacuum energy is tuned to zero. Parameters $c_i$ are defined as
\be
c_i\equiv s_i\frac{\partial \ln\kappa(s_i)}{\partial s_i}\,,\,\,\,\,\,\,{\rm no\,\,sum\,\,over\,\,}i\,,
\ee
and satisfy
\be
\sum_{i=1}^Nc_i=0\,,
\ee
because $\kappa(s_i)$ is a homogeneous function of degree zero. At the minimum of the potential
parameters ${a}_i$ and $c_i$ can be determined by solving a system of $2N$ coupled
transcendental equations
\ba\label{aici}
&& s_i\frac{\partial \hat K}{\partial s_i}\Big |_{_{s_i=\frac 1{N_i}\left({a}_i+c_i\frac{\phi_c^2}3r\right)}}+3{a}_i\,=0\,,\\
 &&s_i\frac{\partial \ln\kappa(s_i)}{\partial s_i}\Big |_{_{s_i=\frac 1{N_i}\left({a}_i+c_i\frac{\phi_c^2}3r\right)}}-c_i\,=0\,.\nonumber
\ea

Once again, the structure of the soft supersymmetry breaking terms  remains virtually unchanged in the leading order in $1/P_{eff}$ expansion
and does not depend on the precise details of the function $\kappa(s_i)$. In fact, the only modification compared to the previously derived expressions
is the replacement of the following combination
\be
\frac {\phi_0^2}{V_X}\rightarrow\kappa(s_i)\frac {\phi_0^2}{V_X}\,,
\ee
whose vev at the minimum is given by (\ref{scf}) and is exactly the same as before! However, it turns out that
the soft supersymmetry breaking terms now become slightly sensitive to the compactification details via the subleading corrections. 
The most sensitive parameter is the tree-level gaugino mass. Up to an overall phase it is given by
\be\label{fgm}
m_{1/2}^{tree}\approx -\frac { m_{3/2}}{P_{eff}}\left(1+\frac {2V_X}{(Q-P)\kappa(s_i)\phi_0^2}+{\cal O}\left(\frac 1{P_{eff}}\right)\right)= -\frac { m_{3/2}}{P_{eff}}\left(1+\frac 2{(Q-P)\phi_c^2}+\frac {\delta}{P_{eff}}\right)\,,
\ee
where we introduced a phenomenological quantity $\delta\sim {\cal O}(1-10)$ in order to parameterize the additional correction.
In the numerical toy examples we studied, we obtain a ${\cal O}(1-10)\%$ variation in the value of the tree-level gaugino mass, while the other
soft terms vary by less than $1\%$. For the sake of completeness we shall list the expressions for the remaining soft breaking terms
\ba
m_a^{AM}&\approx&\frac{\alpha_{GUT}}{4\pi}\left (\left(3C_a-\sum_{\alpha} C_a^{\,\alpha}\right)K_1+\frac 23 K_2\sum_{\alpha}C^{\,\alpha}_a\right )\times m_{3/2}\,,\\
\nonumber\\
{\tilde A}_{\alpha\beta\gamma}^{tree}&\approx&\left(K_2+\frac {4\pi }{P_{eff}}\left(1+\frac{2V_X}{(Q-P)\kappa(s_i)\phi_0^2}\right)V_{Q^{\alpha\beta\gamma}}\right)\times m_{3/2}\,,\nonumber\\
\nonumber\\
{\tilde A}_a^{AM}&\approx&-\frac 1{16\pi^2}\gamma_aK_1\times m_{3/2}\,,\nonumber\\
\nonumber\\
m_{\alpha}^2&\approx&(1-c(s_i))\left(m_{3/2}^2-\frac 73 \left(m_{1/2}^{tree}\right)^2\right)+\lambda\left(m_{1/2}^{tree}\right)^2\,,\nonumber
\ea
where we defined
\ba
&&K_1\equiv1-\frac 13\left(1+\frac{2V_X}{(Q-P)\kappa(s_i)\phi_0^2}\right)\left(\kappa(s_i)\frac{\phi_0^2}{V_X}+\frac 7{P_{eff}}\right)\,,\\\nonumber\\
&&K_2\equiv\left(1+\frac{2V_X}{(Q-P)\kappa(s_i)\phi_0^2}\right)\left(\left(1-c(s_i)\right)\kappa(s_i)\frac{\phi_0^2}{V_X}+\frac{3\lambda}{P_{eff}}\right)\,.\nonumber
\ea
To illustrate the high accuracy of the analytical results presented here we present a simple toy example with two moduli.
\be
V_X=s_1s_2^{4/3}\,,\,\,\,\,\,\kappa(s_i)=1+\frac{s_1^2}{s_2^2}\,.
\ee
For the following choice of the parameters in the superpotential
\be
A_1=27\,,\,\,A_2\approx 2.638\,,\,\,P=27\,,\,\,Q=30\,,\,\,N_1=1\,,\,\,N_2=2\,,
\ee
where $A_2$ was tuned to cancel the leading contribution to the vacuum energy, we obtain numerically
\be\label{snum2}
s_1\approx 71.67\,,\,\,s_2\approx 13.02\,,\,\,\phi_0\approx 7.24\,,\,\,\Rightarrow\,\,\phi_c^2\approx 0.746\,,\,\,P_{eff}\approx 61.68\,.
\ee
Notice that the above values of $\phi_c^2$ and $P_{eff}$ are extremely close to those in (\ref{mvev}) and (\ref{CCcond}),
obtained numerically for the case when $\kappa(s_i)=1$. These values are also in very good agreement with the corresponding analytical
results. By solving the system (\ref{aici}) for the above example we obtain
\be
{a}_1=1\,,\,\,{a}_2=\frac 43\,,\,\,c_1=-c_2\approx 1.939\,,
\ee
and using the above values in the analytic expression (\ref{simod}) with $P_{eff}\approx 61.68$ we find
\be
s_1\approx 72.5\,,\,\,s_2\approx 12.8\,,
\ee
which agree well with the numerically obtained  values in (\ref{snum2}). 
To verify the tree-level gaugino mass formula numerically we need to know the integers $N_1^{\rm vis}$ and $N_2^{\rm vis}$ for
the visible sector gauge kinetic function. Here we list three representative examples where we varied $N_1^{\rm vis}$ and $N_2^{\rm vis}$,
while keeping everything else fixed
\ba
&&N_1^{\rm vis}=1\,,\,\,N_2^{\rm vis}=0\,,\,\,\Rightarrow \,\,m_{1/2}^{tree}\approx -0.029\times m_{3/2}\,,\\
&&N_1^{\rm vis}=1\,,\,\,N_2^{\rm vis}=3\,,\,\,\Rightarrow \,\,m_{1/2}^{tree}\approx -0.032\times m_{3/2}\,,\nonumber\\
&&N_1^{\rm vis}=0\,,\,\,N_2^{\rm vis}=1\,,\,\,\Rightarrow \,\,m_{1/2}^{tree}\approx -0.037\times m_{3/2}\,,\nonumber
\ea
which demonstrate a mild dependence of the tree-level gaugino mass on the compactification-specific details.
Using $\phi_c^2\approx 0.746$ together with $P_{eff}\approx 61.68$ in the analytic formula (\ref{fgm}) for $Q-P=3$
we obtain
\be\label{trgm}
m_{1/2}^{tree}\approx -\left(0.031+0.00026\times\delta\right)\times m_{3/2}\,,
\ee
which is in fairly good agreement with the numerical results. The numerical results for the remaining
soft terms are given by the following expressions
\ba\label{sfttrms}
&&m^{AM}_a\approx \alpha_{GUT}\left(0.0337\left(3C_a-\sum_{\alpha} C_a^{\,\alpha}\right)
+0.0792\left(1-c\right)\sum_{\alpha} C_a^{\,\alpha}\right)\times m_{3/2}\,,\\
\nonumber\\
&&{\tilde A}_a\approx 1.49(1-c)m_{3/2}-\gamma_a\times  0.0027\times m_{3/2}\,\nonumber\\
\nonumber\\
&&m_{\alpha}\approx(1-c)^{1/2}\,m_{3/2}\,,\nonumber\\
\nonumber\\
&&K_1\approx 0.424 \,,\,\,\,\,\,\,\,K_2\approx 1.494\left(1-c\right)\,,\nonumber
\ea
and are virtually unchanged compared to the numerical results in (\ref{mamed}), (\ref{atri2}) and
(\ref{scl}), computed for the case when $\kappa(s_i)=1$. The values obtained from the corresponding analytic expressions
are in good agreement with the numerical values above and give essentially the same results as in the case when 
$\kappa(s_i)=1$ because the values of $P_{eff}$ and $\phi_c^2$ barely changed. Thus, the effect of
including a non-trivial function $\kappa(s_i)$ in the Kahler potential for the hidden sector matter fields can
 be reliably described by a single parameter $\delta$ that appears in the subleading contributions to the 
tree-level gaugino mass, while the remaining soft terms stay essentially unaffected.

\section{Electroweak scale spectrum}
In order to obtain the corresponding MSSM spectrum at the electroweak scale we need to RG-evolve
all the masses and couplings from the GUT scale down to the electroweak scale. This procedure
was described in great detail in \cite{Acharya:2008zi}. Here we will only highlight a few important
points and give the final results. 

As we have seen in the previous section, at the GUT scale, the gaugino masses are non-universal 
and highly suppressed relative to the gravitino mass. On the other hand, unless $c$ is very close to
one, the scalars, trilinear couplings and the $\mu$-term are all of order $m_{3/2}$. Hence, we can define
a scale $m_s$ at which all the heavy states decouple and the effective theory below 
that scale is the Standard Model plus gauginos. More specifically, we can choose the decoupling scale $m_s$ to be
the geometric mean of the stop masses 
\be\label{decscale}
m_s=\sqrt{m_{\tilde t_1}m_{\tilde t_{2}}}\,.
\ee
This is okay as long as the mass differences between the lightest stop and the other heavy states is not too large.
Then, the running can be done at one loop in two stages with tree-level matching at the scale $m_s$.
This method, however, does not capture the two-loop effects, which may give significant contributions
to the running. Thus, in what follows we will utilize the SOFTSUSY package \cite{Allanach:2001kg} and perform the
running at two-loops with the full the MSSM spectrum and account for the effects from the heavy scalars 
via threshold corrections.

\subsection{Gauginos}
As one notices from (\ref{sfttrms}), due to the anomaly mediated contribution,
the gaugino masses are sensitive to the value of $\alpha_{GUT}$. However, the value of $\alpha_{GUT}$
is only determined once we know the exact spectrum and run the gauge coupling up to
the GUT scale. Therefore, there is a feedback mechanism, which allows us to completely
fix the gaugino masses by imposing the gauge coupling unification. In practice, we
first pick an initial  value of $\alpha_{GUT}\sim 1/25$, compute the gaugino masses, scalar
masses, trilinears, etc. at the GUT scale and run them down to the electroweak scale 
where we compute the spectrum. We then run the gauge couplings up using two-loop RGEs
to check if they unify at the same value of $\alpha_{GUT}$ as we chose to compute
the gaugino masses. If there is disagreement, we change the value of $\alpha_{GUT}$ by
a small increment and repeat the steps until there is a match. In addition, parameter $\eta$ which 
appears inside the gaugino masses and was defined in (\ref{eta}) can be safely set to one. 
This is because as one varies the integers $w$ and $q$ inside (\ref{rstorsion}) over a 
reasoble range, the torsion, unless specifically tuned, is so small that that
the KK threshold corrections can be neglected. 

\begin{figure}[h!]
\begin{tabular}{cc}
\leavevmode \epsfxsize 7.7 cm \epsfbox{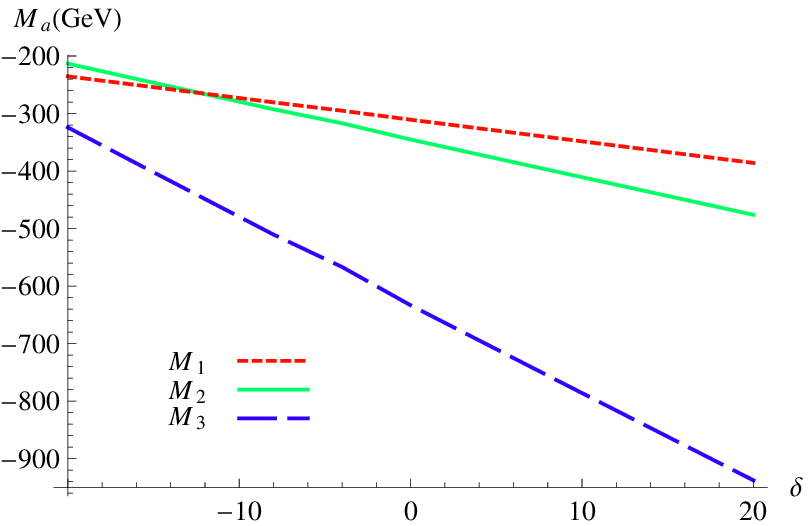}&
\leavevmode \epsfxsize 8.5 cm \epsfbox{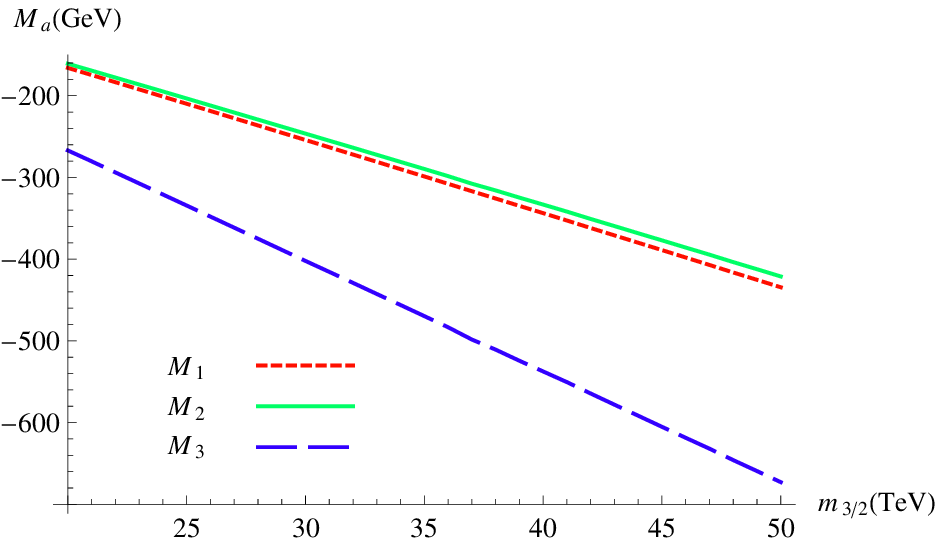}
\end{tabular}
\caption{{\protect\footnotesize Left panel: Gaugino mass parameters at the electroweak scale as functions of $\delta$.  
In the above computation, we used SOFTSUSY with the high scale input $m_{3/2}= 30\,{\rm TeV}$, $\mu<0$, 
$\tan\beta=2.5$, $c=0$.
We have verified by examining the neutralino mixing matrix that at $\delta\approx -12$ the
LSP type changes from Wino to Bino as $\delta$ is increased.\\
Right panel: Gaugino mass parameters at the electroweak scale as functions of $m_{3/2}$.
For each data point we checked the neutralino mixing matrix to confirm that the LSP is Wino-like.
The plot was generated using SOFTSUSY with $\mu<0$, $c=0$, $\delta=-15$, $\tan\beta=2.5$. 
The KK threshold correction to the visible sector gauge kinetic function was neglected in both plots.}}
\label{Fig:gauginoplots}
\end{figure}

Since $M_{Higgsino}\sim\mu\sim {\cal O}(m_{3/2})$, there is a substantial threshold 
contribution from the Higgs-Higgsino loops which has to be taken into account when computing bino 
($M_1$) and wino ($M_2$) masses \cite{Acharya:2008zi}, \cite{Pierce:1996zz,chisplit3,ADG}:
\be\label{correc}
\Delta M_{1\,,2}\approx-\frac{\alpha_{1\,,2}}{4\pi}\frac{\mu\sin(2\beta)}
{\left(1-\frac{\mu^2}{m_A^2}\right)}\ln\frac{\mu^2}{m_A^2}\approx \frac{\alpha_{1\,,2}}{4\pi}\mu=\frac{\alpha_{1\,,2}}{4\pi}Z^1_{eff}m_{3/2}\,.
\ee
In the above expression we expanded the logarithm using $\frac{\mu^2}{m_A^2}\sim 1$ and used 
$\tan\beta\sim{\cal O}(1)$. The latter is especially true when $2\,Z^1_{eff}\approx Z^2_{eff}$. 
We also relied on the fact that the supersymmetric $\mu$-term almost does not change with the 
RG evolution so one can use (\ref{muterm}). 
Since $m_{3/2}\sim{\cal O}(10){\rm TeV}$, the above correction to $M_2$ can be as large as a few hundred GeV. 
It turns out that this contribution is intimately related to the value of parameter $c$ that directly affects the scalar masses
and indirectly forces the value of $\mu$ to get smaller, as the scalars get lighter. For $0\leq c \lesssim 0.05$ and 
$0.8\lesssim \mu /m_{3/2}$ the contribution (\ref{correc}) is actually  large enough
to completely alter the nature of the LSP depending on the sign of $\mu$. In this respect, the sign of $\delta$ that parameterizes 
the subleading corrections to the tree-level gaugino mass (\ref{trgm}) also plays an important role.
In particular, from the left plot in Fig \ref{Fig:gauginoplots} where we picked $c=0$  there is a region where $\delta\lesssim- 12$ 
such that the LSP is Wino-like, while for $\delta\gtrsim -12$ the LSP becomes mostly Bino.
Furthermore, as one can see from the plot, there exists a small range of values where $M_1$ and $M_2$ become nearly degenerate. 
This is certainly an intriguing possibility, which may provide for a well-tempered neutralino candidate \cite{ADG}. Note that in the Wino-like LSP case, 
the lightest chargino and neutralino are degenerate at tree-level, i.e. $\widetilde \chi_1^0=\widetilde \chi_1^{\pm}=M_2$. 
However as we take into account the 1-loop contribution from the gauge bosons \cite{Pierce:1996zz}, 
this degeneracy is removed, as is seen from the corresponding entries in Table II.
 Such splitting was discussed in detail for the pure anomaly mediation scenario in 
\cite{Feng:1999fu,Asai:2007sw} and is given by
\be
\Delta M_{1-loop}=\frac{\alpha_2M_2}{4\pi}\left(f\left(\frac{m_W}{M_2}\right)-c^2_Wf\left(\frac{m_Z}{M_2}\right)-s^2_Wf\left(0\right)\right)\,,
\ee
where $f(a)\equiv\int_0^1 dx(2+2x)\ln[x^2+(1+x)a^2]$. Typically we obtain $160\,{\rm MeV}<\Delta M_{1-loop}<200\,{\rm MeV}$. 
Because of this, the lightest charginos are quasi-stable and decay into LSP plus soft pions or soft leptons. 
In the collider context such decays would  take place well inside the detector leaving  short charged tracks. 

In addition to (\ref{correc}), the EW threshold corrections from gaugino-gauge-boson loops must 
also be included, especially for the gluino
\begin{eqnarray}\label{thgluino}
\Delta M_3^{\mathrm{rad}}=\frac{3\alpha_3}{4\pi}\left(3\ln\left(\frac{M_{\mathrm{EW}}^2}{M_3^2}\right)+5\right)M_3\,.
\end{eqnarray}

Unfortunately, we have no technical handle on the size of the parameter $c$, though we expect it to be
small and hence a Wino LSP is quite generic when $c\approx 0$.
As we increase the value of parameter $c$, the LSP quickly becomes Bino-like. There are two reasons for this
effect. First, the ratio $M_2/M_1$ at the GUT scale grows as $c$ is increased from zero to one.
At the same time, the scalars and higgsinos become lighter relative to the gravitino mass. In particular,
for a fixed $m_{3/2}$ the lower bound on the Higgs mass forces us to consider somewhat larger 
values of $\tan\beta$ which in turn leads to smaller values of $\mu$ thus significantly
reducing the contributions (\ref{correc}) from higgsinos. Recall that it was primarily
due to this contribution from heavy higgsinos for $\mu<0$ that the LSP could become Wino-like.
Thus, due to the increase of $M_2/M_1$ at the GUT scale and the decrease in the higgsino mass, the Wino-like
LSP case becomes rapidly excluded as we increase the value of $c$. Benchmarks 4 and 5 in Table II demonstrate
that for generic values of $c>0$ the LSP is always Bino-like.

\begin{table}[h!]
\label{spectrum1}
\begin{tabular}{||c||c|c|c|c|c|c|c||}
\hline\hline parameter\rule{0pt}{3.0ex}\rule[-1.5ex]{0pt}{0pt} &
Point~1 & Point~2 & Point~3 & Point~4 & Point~5 & Point~6 & Point~7 \\ \hline\hline
$m_{3/2}$ \rule{0pt}{3.0ex} & 20000 & 20000 & 20000 & 20000 & 30000 & 50000 & 30000 \\
$\delta$ \rule{0pt}{3.0ex} & -15 & -12 & 0 & -15 & 15 & -15 & -15\\
$c$ \rule{0pt}{3.0ex} & 0 & 0 & 0 & 0.1 & 0.5 & 0 & 0\\
$\tan\beta$ \rule{0pt}{3.0ex} & 3 & 2.65  & 2.65 & 3 & 3 & 2.5 & 3 \\
\hline
$\mu $ & -11943 & -13377 & -13537 & -10969 & -10490 & -34019 & +17486\\
\hline
${\rm LSP\,\, type} $ & {\rm Wino} & {\rm Wino} & {\rm Bino} & {\rm Bino} & {\rm Bino} & {\rm Wino} & {\rm Bino}\\
\hline
$M_{1}$ & 165 & 173 & 203 & 181 & 484 & 434 & 252 \\
$M_{2}$ & 158 & 173 & 225 & 189 & 662 & 421 & 242 \\
$M_{3}$ & 262 & 297 & 423 & 328 & 1328 & 673 & 395 \\
\hline
$m_{\tilde g}$ & 401 & 449 & 622 & 492 & 1784 & 1001 & 596.8 \\
$m_{\widetilde \chi_1^0}$ & 145.1 & 155.6 & 189 & 170 & 473 & 373.4 & 271 \\
$m_{\widetilde \chi_2^0}$ & 153 & 159 & 214.3 & 181.5 & 702.4 & 397 & 334.2 \\
$m_{\widetilde \chi_3^0}$ & 11905 & 13321 & 13479 & 10938 & 10486 & 33886 & 17441 \\
$m_{\widetilde \chi_4^0}$ & 11906 & 13322 & 13479 & 10939 & 10487 & 33886 & 17442 \\
$m_{\widetilde \chi_1^{\pm}}$ & 145.2 & 155.8 & 214.5 & 181.7 & 702.6 & 373.6 & 334.2 \\
$m_{\widetilde \chi_2^{\pm}}$ & 11970 & 13383 & 13540 & 11001 & 10560 & 34044 & 17540 \\
$m_{\tilde d_L}$,\,$m_{\tilde s_L}$ & 19799 & 19803 & 19809 & 18785  & 21052 & 49524 & 29727 \\
$m_{\tilde u_L}$,\,$m_{\tilde c_L}$ & 19801 & 19812 & 19818 & 18784 & 21034 & 49600 & 29725 \\
$m_{\tilde b_1}$ & 15342 & 15250 & 15224 & 14635 & 16783 & 38473 & 23236 \\
$m_{\tilde t_1}$ & 9130 & 8779 & 8662 & 8928 & 11151 & 22887 & 14264 \\
$m_{\tilde e_L}$,\,$m_{\tilde \mu_L}$ & 19948 & 19948 & 19951 & 18926 & 21164 & 49889 & 29930 \\
$m_{\tilde \nu_{e_L}}$,\,$m_{\tilde \nu_{\mu_L}}$ & 19950 & 19954 & 19952 & 18927 & 21168 & 49903 & 29934 \\
$m_{\tilde \tau_1}$ & 19934 & 19941 & 19940 & 18914 & 21156 & 49874 & 29909 \\
$m_{\tilde \nu_{\tau_L}}$ & 19936 & 19944 & 19942 & 18916 & 21158 & 49876 & 29913 \\
$m_{\tilde d_R}$ & 19848  & 19851 & 19845 & 18832 & 21096 & 49694 & 29794 \\
$m_{\tilde u_R}$,\,$m_{\tilde c_R}$ & 19850  & 19853 & 19858 & 18832 & 21094 & 49700 & 29792 \\
$m_{\tilde s_R}$ & 19849 & 19851 & 19856 & 18832 & 21096 & 49695 & 29767 \\
$m_{\tilde b_2}$ & 19829 & 19833 & 19838 & 18810 & 21075 & 49669 & 29758 \\
$m_{\tilde t_2}$ & 15342 & 15251 & 15224 & 14635 & 16783 & 38470 & 23235 \\
$m_{\tilde e_R}$,\,$m_{\tilde \mu_R}$ & 19978 & 19977 & 19977 & 18953 & 21196 & 49948 & 29966 \\
$m_{\tilde \tau_2}$ & 19948 & 19957 & 19955 & 18930 & 21174 & 49904 & 29928 \\
$m_{h_0}$ & 116.4 & 114.3 & 114.6 & 116.0 & 115.9 & 115.1 & 114.6 \\
$m_{H_0}$,\,$m_{A_0}$,\,$m_{H^{\pm}}$ & 24614 & 25846 & 25943 & 23158 & 25029 & 65690 & 36623 \\
${\tilde A}_t$ & 12159 & 11539 & 11445 & 10898 & 9626 & 30139 & 18812 \\
${\tilde A}_b$ & 27381 & 27321 & 27427 & 24744 & 21850 & 68441 & 41148 \\
${\tilde A}_{\tau}$ & 30068 & 30092 & 30124 & 27109 & 23022 & 75221 & 45099 \\ \hline\hline
\end{tabular}%
\caption{Low scale spectra for seven benchmark $G_2$-MSSM models generated by SOFTSUSY package. 
All masses are in GeV. The top mass was taken to be $m_t=171.3\,{\rm GeV}$. Here we only give absolute values of
the gaugino masses and suppress the relative phases.
The spectra are largely determined by four parameters $m_{3/2}$, $\delta$, $c$ and $\tan\beta$. The Kaluza-Klein threshold corrections
to the gaugino masses have been neglected. For the above spectra, the gauge couplings unify at the value of $\alpha_{GUT}^{-1}\approx 26$ at the scale $M_{GUT}\approx 2\times 10^{16}\,{\rm GeV}$.}
\end{table}

Of course, pure Bino LSP is almost certainly excluded by the standard cosmological considerations \cite{ADG}.
Namely, because binos do not annihilate efficiently, the dark matter relic density
becomes unacceptably large. However, this problem can be avoided when the higgsinos, which annihilate efficiently,
are light enough to mix with gauginos. If the higgsino component of the LSP is significant, it can easily reduce
the relic density to acceptable levels by increasing the annihilation crossection of the LSPs.
It turns out that for generic values of $c$, $0\leq c<1$, the higgsinos are always much heavier
than the gauginos. This is because at the decoupling scale $m_s$, the $\mu^2$-term must be of the
same order of magnitude as $m_{H_u}^2$ to give a correct value of the Z-boson mass, and 
since for typical values of $c$ we get $|m_{H_u}^2|>>M_{1,\,2}^2$, the higgsinos do not mix with gauginos.

\subsection{Squarks and sleptons}
Recall that at the GUT scale all the squarks and sleptons have a universal mass (\ref{scl}), which for generic
values of $c$ ($0\leq c<1$) is smaller but nevertheless typically of the same order of magnitude as the gravitino mass. 
However, as we evolve these down to the electroweak scale, the third generation scalars  become significantly lighter 
whereas the first and second generation scalars experience a very mild change in their masses.  
Indeed, because the third generation Yukawa couplings are large, the stops, 
sbottoms, and staus are affected through the corresponding trilinear couplings (\ref{atri2}), which are of ${\cal O}(m_{3/2})$. 
As one can see from Table II, this effect is especially dramatic for the lightest stop $\tilde t_1$. 
Yet, it is still much heavier than the gauginos and is effectively decoupled from the spectrum 
at the electroweak scale.

However, since gluinos can be pair produced at the LHC via gluon fusion, the gluinos (which have to
decay via a  quark-squark pair) have a sizeable branching fraction into top-stop -- precisely because
the stop is the lightest squark. This leads to events containing up to four top quarks at the LHC \cite{multitop}.

\subsection{Radiative electroweak symmetry breaking}\label{ewsb}
The existence of the electroweak symmetry breaking (EWSB) in the effective theory below the decoupling scale 
$m_s$ is determined by whether there exists a negative eigenvalue in the Higgs mass matrix
\begin{eqnarray}\label{HiggsMatrix}
\left(
\begin{array}{cc}
m_{H_u}^2+\mu^2 & -B\mu \\
-B\mu & m_{H_d}^2+\mu^2%
\end{array}
\right)  \label{Eq-higgsmass}
\end{eqnarray}
at the scale where the scalars decouple \cite{DG}. Recall that at the GUT scale, $m_{H_u}^2=m_{H_d}^2=(1-c)\,m_{3/2}^2$ 
whereas $\mu=Z^{\,1}_{eff}m_{3/2}$ and $B\mu=Z^{\,2}_{eff}m_{3/2}^2$. It is well known that the positive contribution into the running of 
the up Higgs mass parameter squared from the stop is crucial for radiative EWSB as it drives $m_{H_u}^2$ negative.  It turns out that for 
a fixed value of the gravitino mass $m_{3/2}$, as we vary parameter $c$ 
there exists a narrow range of values $Z^{\,1,\,2}_{eff}$ for which the matrix (\ref{HiggsMatrix}) has a 
negative eigenvalue above the decoupling scale $m_s$ defined in (\ref{decscale}).

However, unless we force $Z^{\,2}_{eff}<< 2\,Z^{\,1}_{eff}$, all the entries in the above matrix are ${\cal O}(m^2_{3/2})$. 
Therefore, both the lightest Higgs mass and the Z-boson mass 
naturally come out to be of ${\cal O}(m_{3/2})$. For the same reason, there is little mixing between the two 
Higgs doublets and naturally $\tan\beta\sim{\cal O}(1)$ is predicted in our framework. 
In practice, parameters $Z^{\,1,\,2}_{eff}$ must be tuned in such a way that the corresponding eigenvalue 
turns negative right at the decoupling scale \cite{DG} so that $m_Z\approx 91\,{\rm GeV}$. 
This fine tuning is a manifestation of the so-called {\it little hierarchy problem} - the hierarchy between the 
electroweak scale $M_{EW}\sim{\cal O}(100)\,{GeV}$ and the scale where the scalars 
decouple $m_s\sim{\cal O}(10)\,{\rm TeV}$. Once $M_Z$ is tuned, 
the Standard Model Higgs mass turns out to be $m_h<130\,{\rm GeV}$.

\acknowledgments We would like to thank Jacob Bourjaily, Volker Braun, Kiwoon Choi, Joseph Conlon, Phill Grajek,
Shamit Kachru, David Morrissey, Brent Nelson, Jogesh Pati, Aaron Pierce, Stuart Raby, 
Timo Weigand and Alexander Westphal for useful discussions and in particular acknowledge the input of our collaborators
Gordy Kane, Piyush Kumar, Jing Shao and Scott Watson without whom this work would not
have been possible. BSA thanks the MCTP Ann Arbor for hospitality and support when this 
project was initiated. The research of KB is supported in part by the US Department of Energy. 
KB thanks SLAC for their hospitality.

\end{document}